\documentclass[fleqn,usenatbib]{mnras}

\usepackage[T1]{fontenc}

\DeclareRobustCommand{\VAN}[3]{#2}
\let\VANthebibliography\thebibliography
\def\thebibliography{\DeclareRobustCommand{\VAN}[3]{##3}\VANthebibliography}

\usepackage{graphicx}	
\usepackage{amsmath}	
\usepackage{amssymb}	
\usepackage{newtxtext,newtxmath}
\usepackage{multirow}
\usepackage{lscape}
\usepackage{makecell}
\usepackage{subfigure}
\usepackage{float}
\usepackage{newtxtext,newtxmath}
\usepackage{lineno}

\usepackage{appendix}

\makeatletter
\newcommand{\labtext}[2]{%
  \@bsphack
  \csname phantomsection\endcsname 
  \def\@currentlabel{#1}{\label{#2}}%
  \@esphack
}
\makeatother

\title[MWL Analysis of Blazars with High-Significance Periodicity]{Multiwavelength Analysis of \textit{Fermi}-LAT Blazars with High-Significance Periodicity:  Detection of a Long-Term Rising Emission in PG~1553+113}

\author[P. Pe\~nil et al.]{
P. Pe\~nil,$^{1}$\thanks{E-mail: ppenil@clemson.edu}
J.R. Westernacher-Schneider,$^{1,2}$
M. Ajello, $^{1}$\thanks{E-mail:majello@clemson.edu}
A. Dom\'inguez,$^{3}$
S. Buson,$^{4}$
J. Otero-Santos,$^{5,6,7}$
\newauthor
L. Marcotulli,$^{8}$\thanks{NHFP Einstein Fellow}
N. Torres$-$Alb\`a,$^{1}$
J. Zrake,$^{1}$
\\
$^{1}$Department of Physics and Astronomy, Clemson University, Kinard Lab of Physics, Clemson, SC 29634-0978, USA\\
$^{2}$Leiden Observatory, Leiden University, P.O. Box 9513, 2300 RA Leiden,
The Netherlands\\
$^{3}$IPARCOS and Department of EMFTEL, Universidad Complutense de Madrid, E-28040 Madrid, Spain\\
$^{4}$Julius-Maximilians-Universit\"at, 97070, W\"urzburg, Germany\\
$^{5}$Instituto de Astrof\'isica de Canarias (IAC), E-38200 La Laguna, Tenerife, Spain\\
$^{6}$Universidad de La Laguna (ULL), Departamento de Astrof\'isica, E-38206 La Laguna, Tenerife, Spain\\
$^{7}$Instituto de Astrof\'isica de Andalucía (CSIC), Glorieta de la Astronomía s/n, 18008 Granada, Spain\\
$^{8}$ Department of Physics, Yale University, 52 Hillhouse Avenue, New Haven, CT 06511, USA\\
}

\date{Accepted 2023 October 18. Received 2023 October 11; in original form 2023 February 28}

\pubyear{2023}

\begin{document}
\label{firstpage}
\pagerange{\pageref{firstpage}--\pageref{lastpage}}
\maketitle

\begin{abstract}
Blazars display variable emission across the entire electromagnetic spectrum, with timescales that can range from a few minutes to several years. Our recent work has shown that a sample of five blazars exhibit hints of periodicity with a global significance $\gtrsim2\,\sigma$ at $\gamma$-ray energies, in the range of 0.1~GeV$<$E$<$800~GeV. In this work, we study their multiwavelength (MWL) emission, covering the X-ray, ultraviolet, optical, and radio bands. We show that three of these blazars present similar periodic patterns in the optical and radio bands. Additionally, fluxes in the different bands of the five blazars are correlated, suggesting a co-spatial origin. Moreover, we detect a long-term ($\approx$10 year) rising trend in the light curves of PG~1553+113, and we use it to infer possible constraints on the binary black hole hypothesis.
\end{abstract}


\section{Introduction} \label{sec:intro}

A blazar is an active galactic nucleus (AGN) with a supermassive black hole (SMBH) launching a relativistic jet towards our line of sight \citep[e.g.][]{ulrich_variability}. Blazar emission is highly variable, spanning the entire electromagnetic spectrum \citep{urry_multiwavelengh} and a wide range of timescales, from minutes to years \citep{urry_variability}. Some blazars have periodic multiwavelength (MWL) emission \citep[e.g.][]{ackermann_pg1553}. However, this temporal behavior is usually interpreted as stochastic \citep[e.g.,][]{covino2019}. 

Complementary to periodicity analysis are temporal correlations among the MWL light curves (LCs), which are used to measure time lags between bands. This information helps to constrain the location of the emission region(s), disentangle which jet models are favored, and understand the variability mechanisms \citep[e.g.][]{cor_kait_bigsample, cor_kait_ovro_bigsample}.

In \citet{penil_2022} (P22, hereafter), we reanalyzed 24 blazars exhibiting possible periodicity in \textit{Fermi}-Large Area Telescope \citep[LAT,][]{fermi_lat} observations, which were originally presented, as part of a larger sample, in \citet{penil_2020} (P20, hereafter). From this sample, we selected the five blazars that exhibited the most significant periodicity in $\gamma$-ray ($\geq$4.0$\sigma$ of local significance, P22). In this study, we aim to characterize the MWL behavior of these five sources, from radio up to $\gamma$-rays, and to infer its cause. Employing the same analysis pipeline presented in P22, we search for similar periodic behavior at longer wavelengths. We further perform MWL LC correlation studies and use complementary tools such as fractional variability and structure function to characterize this MWL variability. This paper is the second in a pair of publications. The first publication (hereafter Paper I) focuses on the other 18 blazars from P22 cataloged as low-significance. In Paper I, we employ the same databases and perform the same variability analysis used in this paper. Regarding periodicity, we used the same methodology as this paper, obtaining 2 sources with a period with $\geq$3.0$\sigma$ (pre-trial). Moreover, the correlation analysis reveals a high correlation between the $\gamma$-ray, optical, and IR bands with delays $<28$ days and the radio band with typical delays of a few hundred days. The structure function shows variability timescales compatible with the periods associated with the emission and shorter timescales, supporting that such variability could originate from instabilities or fluctuations in the accretion disc.

The paper is structured as follows. In Section~\ref{sec:sample}, we introduce our sample of sources and discuss the data collected from \textit{Fermi}-LAT and MWL sources. Section~\ref{sec:methodology} provides a description of the methodology and analysis methods. In Section~\ref{sec:periodicity_correlation_results}, we present our results. Section~\ref{sec:trends} is on the discovery and characterization of new types of variability in PG 1553+113 and PKS 2155$-$304. Section~\ref{sec:discussion} offers an interpretation of the origin of the variability observed in PG 1553+113. In Section~\ref{sec:summary}, we summarize the main results and conclusions drawn from our study. Additionally, there is an Appendix that presents all the blazars MWL LCs and provides supplementary information on our variability studies.

\section{Blazar sample and Data} \label{sec:sample}

\subsection{Source Selection}
In this work, we consider five blazars with periods, T, detected in their high-energy (HE, 0.1~GeV$<$E$<$800~GeV) $\gamma$-ray emission with global significance ($\geq$2.0$\sigma$, P22). This sample was obtained from the periodicity analysis of 12 years of \textit{Fermi}-LAT observations, from August 2008 to December 2020. These blazars are listed in Table \ref{tab:candidates_list}, and their MWL emission are shown in Appendix \ref{sec:plots}.

\begin{table}
\centering
\caption{Our sample of blazars given by their association name, AGN type, redshift, period (in years), local significance, and global significance obtained by P22. The AGN types are flat-spectrum radio quasar (fsrq) and BL Lacertae (bll).}
\label{tab:candidates_list}
\resizebox{\columnwidth}{!}{%
\begin{tabular}{ccccccccc}
\hline
\hline
		Association Name & Type & Redshift & Period & Local (S/N) & Global (S/N) \\
        &  &  & [yr] & [yr] &  &  \\
        \hline
        PKS 0454$-$234 & fsrq & 1.000 & 3.6$\pm$0.4 & 4.1$\sigma$ & 2.0$\sigma$ \\
        S5 0716+714 & bll & 0.127 & 2.7$\pm$0.4 & 4.0$\sigma$ & 2.0$\sigma$ \\
        OJ 014 & bll & 1.148 & 4.2$\pm$0.6 & 4.1$\sigma$ & 2.0$\sigma$ \\
        PG 1553+113 & bll & 0.433 & 2.2$\pm$0.1 & 5.0$\sigma$ & 3.0$\sigma$ \\
        PKS 2155$-$304 & bll & 0.116 & 1.7$\pm$0.1 & 4.1$\sigma$ & 2.0$\sigma$ \\
\hline
\hline
\end{tabular}%
}
\end{table}

\subsection{Multiwavelength Archival Data} \label{sec:wave_data}
We use data from several observatories covering a broad swath of the electromagnetic spectrum. For the X-ray and UV bands, we employ data from the {\it Swift}$/$XRT) \footnote{\url{https://www.swift.ac.uk/analysis/xrt/}}.
The raw count-rate data were taken from the automatic processing of \cite{stroh_monitoring}\footnote{\url{http://www.swift.psu.edu/monitoring/}}. Hardness ratio estimations (HR, ratio between soft (0.3$-$2.0 keV) and hard (2.0$-$10.0 keV) X-rays), also by \cite{stroh_monitoring}, were used in conjunction with the {\it Swift}/XRT detector response to estimate a photon index, $\alpha$, under the assumption that blazar emission in X-rays can be represented by a simple power-law (PL) A(E)=KE$^{-\alpha}$ \citep[][]{ghisellini_canonical_blazars, middei_x_ray_cat}. The count rate and HR were then used to estimate a flux at each given epoch. We use data from \textit{Swift}-UVOT (Ultraviolet and Optical Telescope)\footnote{\url{https://www.swift.ac.uk/about/instruments.php}} for the filters `uvw2' (1928~\AA), `uvm2' (2246~\AA) and `uvw1' (2600~\AA)\footnote{\url{https://www.swift.ac.uk/analysis/uvot/filters.php}}. We perform a data reduction of all available {\it Swift}-UVOT archival observations to provide the optical-to-UV LCs via the standard pipeline, detailed in \citet{poole_swift_analysis}. For all UVOT filters, the source regions are selected as circles of 5$\arcsec$, centered on the source. The background is defined as a circle of 30$\arcsec$ away from the source to avoid contamination. The task \texttt{uvotsource} is employed to extract the magnitudes, which are then corrected for Galactic extinction according to the recommendations in \citet{roming_uvot_analysis}. Finally, the fluxes are derived using the standard zero points listed in \citet{breeveld_calibration_swift}. 

For the optical bands, data from KAIT (Katzman Automatic Imaging Telescope, R-band)\footnote{\url{http://herculesii.astro.berkeley.edu/kait/agn/}}, CSS (Catalina Sky Survey, V-band)\footnote{\url{http://nesssi.cacr.caltech.edu/DataRelease/}}, ASAS-SN (All-Sky Automated Survey for Supernovae, V-band)\footnote{\url{http://www.astronomy.ohio-state.edu/asassn/index.shtml}}, and Tuorla \citep[R-band,][]{tuorla_data}\footnote{\url{http://users.utu.fi/kani/1m/}} are used. We also employ the optical V- and R-bands and near-infrared (IR) I-band data of the American Association of Variable Star Observers (AAVSO)\footnote{\url{https://www.aavso.org/data-download/}}. Data from SMARTS \citep[Small and Moderate Aperture Research Telescope System,][]{Bonning_smarts_paper}\footnote{\url{http://www.astro.yale.edu/smarts/glast/home.php}} in the optical B-, V-, and R-bands, and the near-IR J- and K-bands are also used. The Steward Observatory\footnote{\url{http://james.as.arizona.edu/~psmith/Fermi/}} provides public optical data in V- and R-bands, with photometric and polarimetric observations obtained approximately simultaneously.

We have combined the V- and R-band data from the optical observatories, and the combinations are herein denoted as ``V-band'' and ``R-band.'' In some cases, non-calibrated V- and R-band data from the Steward Observatory are used, which are not combined with data from other observatories. We denote this data as ``Steward-V'' and ``Steward-R'' in the figures of Appendix \ref{sec:plots}. 

Finally, 15 GHz data from OVRO (the Owens Valley Radio Observatory 40-m radio telescope) are used. OVRO has been engaged in a blazar monitoring program supporting the \textit{Fermi} satellite \citep{ovro_monitoring}\footnote{\url{https://sites.astro.caltech.edu/ovroblazars/}}. These data extend for longer than 12 years, to June 2020, when monitoring ceased, and the data were made publicly available.

For all the MWL data, we use the same $\gamma$-ray LCs and 28-day binning as in P22, where the {\it Fermi}-LAT data analysis is described. We search for long-term periodicity ($\sim$years) in the range of [1-6] years. 

\section{Methodology} \label{sec:methodology}
We apply the pipeline described in P22 to all data sets for the periodicity search. The pipeline includes methods that can manage the gaps: Lomb-Scargle \citep[LSP,][]{lomb_1976, scargle_1982}, Generalized LSP \citep[GLSP,][]{lomb_1976, scargle_1982}, Phase Dispersion Minimization \citep[PDM,][]{pdm},  Weighted Wavelet Z-transform \citep[WWZ,][]{wwz},  Enhanced Discrete Fourier Transform with Welch's method \citep[DFT-Welch,][]{welch}, and Markov Chain Monte Carlo Sinusoidal Fitting (MCMC Sine, see P20). These methods can provide different results depending on different factors, independently if they are based on the same algorithm. The performance of such methods depends on the sensitivity to the gaps in the LC (see P20), the use or not of the uncertainties of the data to search for the periodicity (e.g., GLSP), and the sensitivity to noise (see P22).

Analyzing non-stationary time series can be challenging \cite[][]{feigelson_2022}. Specifically, a non-stationary time series is characterized in the frequency domain through its time-varying power spectrum \citep[][]{vaughan_fractional_variability}. This implies that estimating the power spectral density (PSD) cannot be applied uniformly to the entire time series. In such instances, the underlying physical process driving the variability changes over time, resulting in variations in properties such as the PSD \citep[][]{vaughan_2013}. These random fluctuations, typical of a red noise process, do not yield meaningful physical insights \citep[][]{vaughan_fractional_variability}. 

One method to ensure stationarity in a time series is detrending. Detrending is a recommended preprocessing step \citep[e.g.,][]{detrend_welsh} since a trend can introduce contamination in the low-frequency components, potentially leading to the detection of false periodicities in the time series \citep[][]{mcquillan_trend_fake_detection}. This helps in avoiding erroneous results associated with any underlying trends in the data. Nonetheless, detrending can introduce correlations in the data, particularly when data transformations are involved, such as the differencing method, which can introduce autocorrelations if the differenced series still retains some underlying patterns. In our case, we employ linear detrending. Linear detrending can have the effect of seemingly increasing noise correlation in the detrended data, even though it does significantly reduce autocorrelation in the time series. This noise correlation can happen if the linear component is not adequately removed or if the noise in the original data contains some non-random systematic patterns or fluctuations. However, in our specific case, we can rule out the latter scenario, as blazar LCs are known for exhibiting red noise characteristics \citep[e.g.,][]{vaughan_fractional_variability}.

In our analyses, the methods of our methodology are effective in managing non-stationary LCs. Moreover, we take steps to verify the stationary nature of the data, post-detrending, to ensure that any previous spurious effects associated with non-stationary time series are eliminated. To accomplish this, we utilize the augmented Dickey-Fuller test \citep[][]{dickey_fuller} for validation of the stationarity of our data \citep[][]{feigelson_arima}.

We compute the cross-correlations of all the MWL data with the corresponding \textit{Fermi}-LAT LC for each source. The correlations are performed by the \textit{z}-transformed discrete correlation function \citep[\textit{z}-DCF,~][]{zdfc_alexander}. We compute the cross-correlation of contemporary high-flux emission states selected with a Bayesian block algorithm \citep{scargle_bayesian_blocks}. 

The PSD, typically described as a power law, quantifies the variation of a time series as a function of frequency, which relates to the physical origin of the emission \citep{abdo_variability}. We estimate its index with the Power Spectrum Response Method \citep[\texttt{PSRESP},~][]{psresp_uttley}.
Based on the PSD model, we employ the method based on simulating LCs \citep[][]{emma_lc} to estimate the local significance (pre-trial correction, see $\S$\ref{sec:trials}). These simulated LCs exhibit identical PSD and probability density function (PDF) characteristics as the original LC, implying that they share the same noise properties as the original LCs. Additionally, these LCs undergo the same detrending process as the original ones. Consequently, the significance is appropriately calibrated for detrending across all the methods we employ. We use the implementation from \citet{connolly_code} for simulating 20,000 non-periodic LCs. 

\subsection{Global significance} \label{sec:trials}
We correct the local significance inferred from the different methods since we did not have prior knowledge of the periods of the potential signal in our periodicity analysis. This correction is implemented using the look-elsewhere effect to estimate the ``global significance'' \citep[][]{gross_vitells_trial}. We evaluate the ``global significance'' by applying the trial factor to the local significance of each periodicity. The trial factor combines the number of independent periods we search for periodicity, 35, and the size of our sample of blazars, 351 (more details in P22). This yields:
\begin{enumerate}
	\item $\approx$3.5$\sigma$ for a local significance of $5.5\sigma$ 
	\item $\approx$2.8$\sigma$ for local significance of $\approx$5$\sigma$
	\item $\approx$1.8$\sigma$ for a local significance of $\approx$4.5$\sigma$
	\item $<$1$\sigma$ for a local significance $<$4.5$\sigma$.
\end{enumerate}

\section{Periodicity and correlation results}\label{sec:periodicity_correlation_results}

\subsection{Periodicity} \label{sec:periodicty_results}
No relevant results are found for PKS 0454$-$234, S5 0716+714, OJ 014 (see Appendix \ref{sec:periodicity_appendix}). The results of the periodicity analysis for PG 1553+113 and PKS 2155$-$304 are shown in Table \ref{tab:methods_results}. 

For PG~1553$+$113 (see Figure \ref{fig:lc_pg_1553}), we find a periodicity of $\approx$2.2\footnote{All quantities referred to in the text are given without the associated uncertainties to make the text more readable. However, all reported numbers can be found in their respective tables, with the associated uncertainties. We refer the reader to those for details} yr in its UV, optical, and radio emission with local significance $>$2$\sigma$, $\geq$5$\sigma$, and $\geq$5$\sigma$, respectively. Periods with local significance $<$3$\sigma$ are considered non-significant in this work. These results are compatible with the observed periodicity in $\gamma$-rays and consistent with \citet{ackermann_pg1553}, who reported a $\approx$2.2-yr period in $\gamma$-ray, optical and radio bands. In the X-ray band, we find a period of $\approx$1.5 yr with local significance $\approx$2$\sigma$. A secondary period compatible with the finding by \citet{huang_pg1553_xperiod}, of $\approx$2.2 yr, is inferred by LSP and PDM.

Finally, for PKS 2155$-$304 (see Figure \ref{fig:mwl_pks2155}), we find a period of $\approx$1.7 yr for the X-ray band with local significance $\approx$2.5$\sigma$. Additionally, we find a period of $\approx$1.7 yr in the UV and optical bands, a period compatible with that observed in $\gamma$-rays, but not significant ($\leq$2.5$\sigma$ of local significance). However, our result is incompatible with the period reported by \cite{sandrinelli_redfit}, who reported a period of $\approx$0.9 yr. \citet{bhatta_s5_0716} reported a period of $\approx$1.7 yr for the optical band.  

The local significance of the results typically ranges from 2$\sigma$ to 5$\sigma$, depending on the method. Even for the same dataset, the results in terms of period and significance can differ. The disparity in outcomes can be due to several factors. The shorter time coverage and uneven sampling of the MWL data sets also lead to much larger errors of the derived periods than those from the \textit{Fermi}-LAT LCs. Specifically, the methods are affected differently by the gaps in the LCs (see P20 for the complete study). For instance, LSP is the most robust against the missing data, and DFT is the most affected method (with differences of 50\% in detecting period and significance). Another factor is the impact of red noise. Specifically, each method has a different bias to red noise. For instance, the most robust is the GLSP, and the most sensitive is the DFT, with differences in the detection capacity of 50\% (see P22). For this reason, some methods tend to have incompatible periodicities and significance for the same dataset (e.g., S5 0716+714 in the UV band, see $\S$\ref{sec:appendix}).

\subsection{Correlation} \label{sec:correlation_results}
Correlation results of PG 1553+113 and PKS 2155-$304$ are shown in Table \ref{tab:cross_correlation}, including the periods inferred with the \textit{z}-DCF. The correlation results of PKS 0454$-$234, S5 0716+71, and OJ 014 are included in Appendix \ref{sec:correlations_results_appendix}. All delays $<\pm$28 days are compatible with zero lag due to the 28-day binning we use for the \textit{Fermi}-LAT LCs. Negative lags imply that the $\gamma$-rays are leading the MWL emission (i.e.~the $\gamma$-ray flare precedes the other wavelengths), and vice versa.

\begin{figure}	        
    \includegraphics[width=\columnwidth]{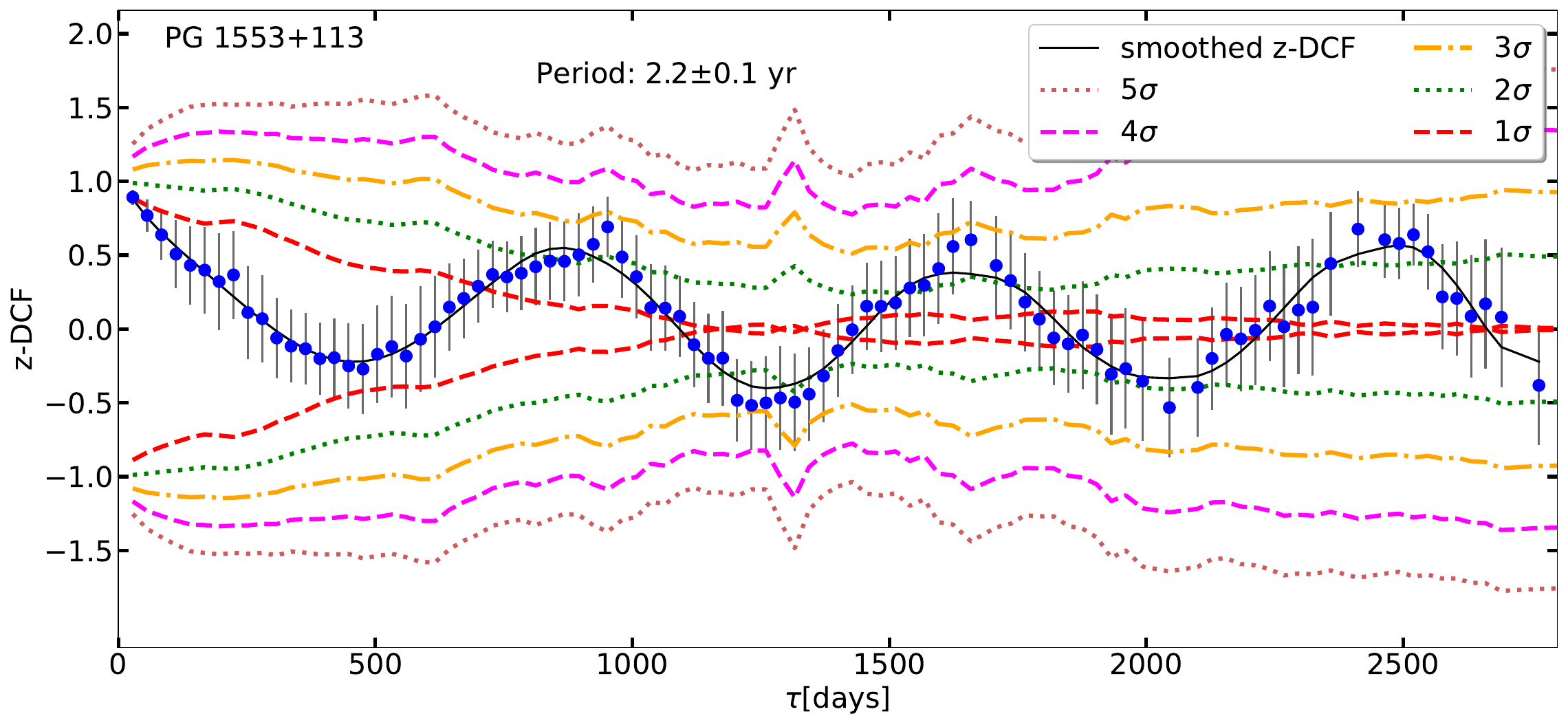}
    \caption{\textit{z}-DCF autocorrelation of PG~1553+113 (V band). The figure shows a period of $\approx$2.2 yr. $\tau$ denotes the time lag in days.}
    \label{fig:autocorrelation}
\end{figure}

Several studies have looked at MWL correlations for PG 1553+113. \cite{cor_kait_ovro_bigsample} find time lags of $\approx$10 days with $\approx$2$\sigma$ (significance before applying any trial correction, as for the other significance reported in this section) between \textit{Fermi}-LAT and KAIT, and $\approx$100 days ($\approx$2$\sigma$) between \textit{Fermi}-LAT and OVRO. In \cite{ackermann_pg1553}, a zero time lag is observed for the optical band and $\approx-$100 days for radio. A negative time lag denotes that the $\gamma$-ray emission is leading the radio. A time lag in the radio band of $-$530 days is claimed by \citet{cor_ovro_bigsample}. In this work, we do not obtain any time lag between the $\gamma$-ray, X-ray, and optical bands (with $>$3$\sigma$), with a delay of radio of $\approx-$200 days ($>$2$\sigma$). The autocorrelation shows a period in optical and radio of $\approx$2.2 yr, compatible with $\gamma$-rays, but not significant (see Figure \ref{fig:autocorrelation}). 

Finally, for PKS 2155$-$304, \citet{cor_v_sample} report zero lag in the optical band, in agreement with our results. We also find zero lag correlation with the X-ray band. In the autocorrelation, we obtain a compatible period in X-ray band with respect to the one inferred in $\gamma$-rays, but it is not significant. Similar results are obtained in the optical band.

\subsubsection{Correlation results between X-rays and $\gamma$-rays} 

A lack of correlation between X-rays and $\gamma$-rays would signify that different regions are involved in the origin of both emissions. However, the results of Table \ref{tab:cross_correlation} show correlations approximately compatible with zero lag between $\gamma$-rays and X-rays (all lags $<\pm$28 days are compatible with 0 lag due to the 28-day binning of the \textit{Fermi}-LAT). These results may imply a common origin of both emissions, considering the uncertainties due to the 28-day sampling. \citet{sikora_x_ray} discuss two emission scenarios that could help in understanding the disparity in the X-ray and $\gamma$-ray correlations in FSRQs. If the X-ray emission produced by the synchrotron self Compton (SSC) and generated at $d>10$ pc dominates over the X-ray emission produced in the BLR, no correlation is expected. On the other hand, a correlation between X-rays and $\gamma$-rays (and, thus, a co-spatial origin) is possible when the Comptonization of hot dust radiation produces the external Compton (EC), dominating the X-ray emission. This takes place at a parsec scale. Regarding the BL Lac objects, the 0-lag correlation of our results may support the model that the $\gamma$-ray and X-ray emissions are likely produced by the same population of relativistic electrons through synchrotron and SSC processes, respectively \citep[e.g.,][]{abdo_bllac_corr}.

\section{Long-Term Trends}\label{sec:trends}

\subsection{Long-Term Trend in the MWL Emissions of PG 1553+113} \label{sec:trend}
\begin{figure}
\includegraphics[width=\columnwidth]{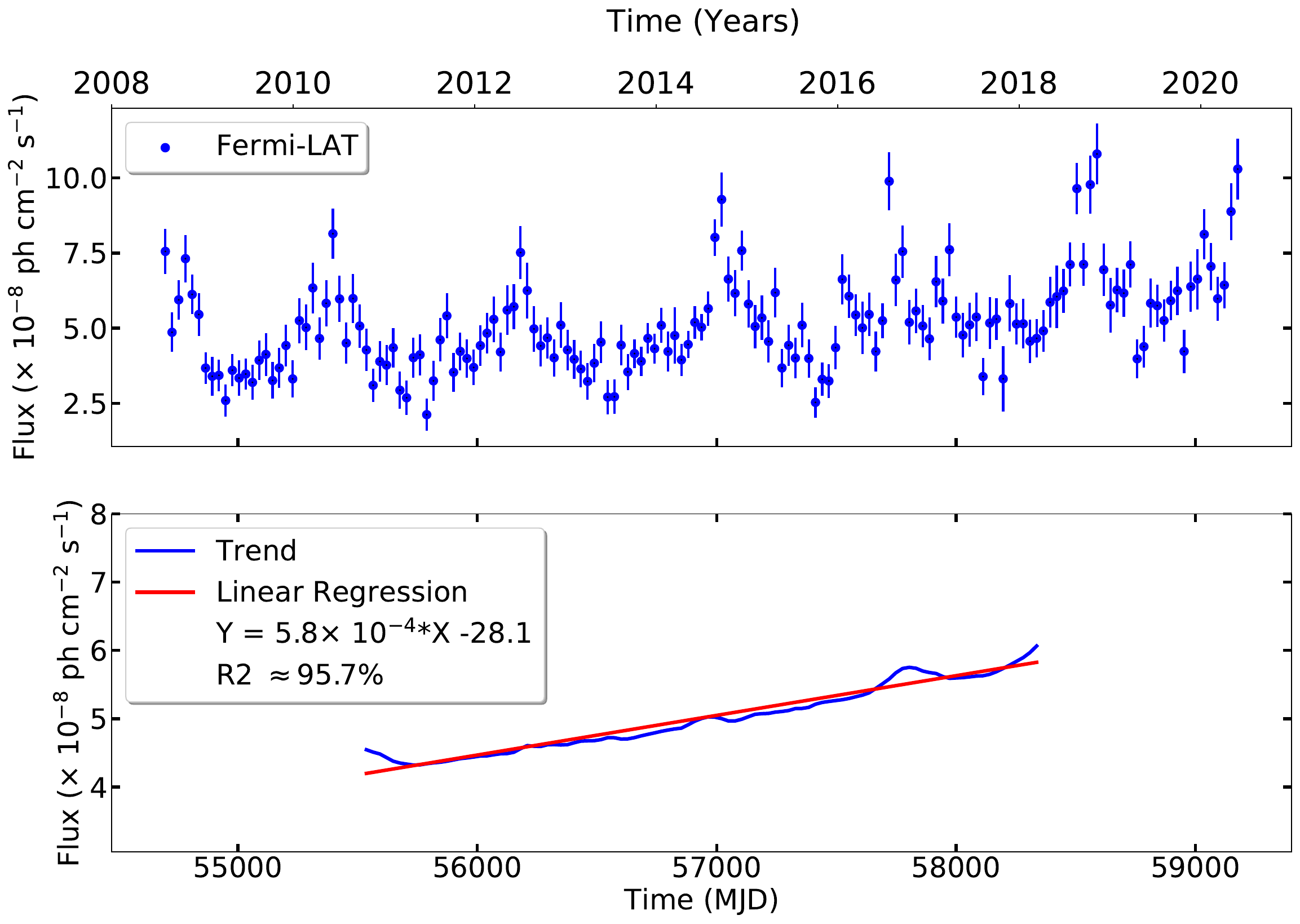}
\caption{Trend decomposition of the $\gamma$-ray emission of PG~1553+113. R$^{2}$ is the R-squared criterion to measure the goodness of the fit.}
\label{fig:trend_1}
\end{figure}

\begin{figure}
\includegraphics[width=\columnwidth]{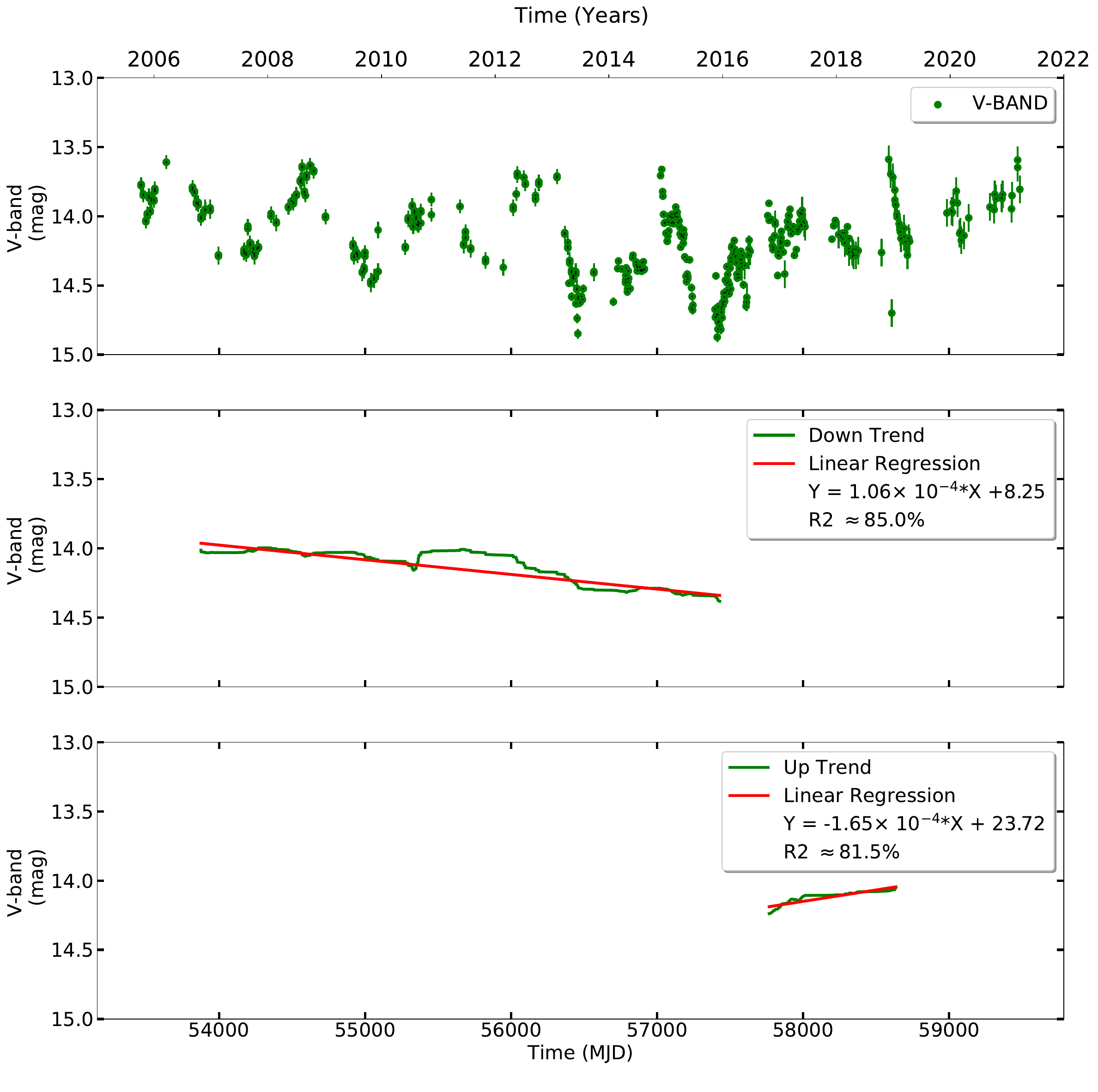}
\caption{Long-term trend decomposition of the V-band emission of PG~1553+113.}
\label{fig:trend_2}
\end{figure}

Observing the MWL bands of Figure \ref{fig:lc_pg_1553}, we can see an increase of the flux (trend) with time in PG~1553+113 in the $\gamma$-ray \citep[][]{rueda_pg_trend}, UV, optical, and radio bands. We estimate each trend by fitting the corresponding LC to a first-degree polynomial function\footnote{Using the \textit{Python} package \texttt{Statsmodels}}. The R-squared (R$^{2}$) criterion is used to measure the goodness of the fit. We consider a fit reliable when R$^{2}\geqslant$75\% \citep{hair_r2}. Examples of this analysis are shown in Figure \ref{fig:trend_1} and Figure \ref{fig:trend_2}. The different slopes inferred from the trend analysis for each waveband have compatible values (with a slope $\approx$2$\times$10$^{-4}$). We also estimate the average amplitudes of the oscillations in the different bands. Firstly, we normalize the LC with the maximum flux of each LC. Then, we measure amplitudes from the consecutive peak-minimum flux in the different bands. Then, we obtain the averaged amplitude. The normalized amplitudes present compatible values within uncertainties: 1.4$\pm$0.2 for $\gamma$ rays, 1.4$\pm$0.4 for X rays, 1.5$\pm$0.2 in the optical band, and 1.5$\pm$0.1 for radio. Finally, our results indicate that the $\gamma$-ray flux ($\approx$2.5$\times$10$^{-8}$ ph cm$^{-2}$ s$^{-1}$) and radio ($\approx$0.15 Jy) begin to increase around the same time, between the years 2011--2013, but in X-ray, UV, and optical bands, the minimum is observed around 2014--2015.

\subsection{Decreasing Long-Term Trend in the MWL emissions of PKS 2155$-$304} \label{sec:trend_pks}
A trend is also observed in PKS 2155$-$304, but, in this case, it is a decrease in the flux \citep[observed in the optical band in ][]{Zhang_pks_trend}. We perform the same trend fit presented in $\S$\ref{sec:trend}. The slopes of the $\gamma$-ray and X-ray bands have comparable values of $\approx$$-$3$\times$10$^{-3}$. Regarding the optical band, the slope inferred is $\approx$$-$2$\times$10$^{-4}$. The normalized amplitudes are comparable values: 1.3$\pm$0.1 for $\gamma$-ray, 0.8$\pm$0.3 for X-ray, and 1.2$\pm$0.1 in the optical band. Finally, the results show that the end of the flux decrease occurs at approximately the same time, around 2014, in all the wavebands. 

\section{The binary hypothesis for PG 1553+113} \label{sec:discussion}
Trends can be associated with red noise, which can mimic periodicity \citep[e.g.][]{vaughan_criticism} or lead to the detection of false periodicity in a time series \citep[][]{mcquillan_trend_fake_detection}. Nonetheless, it is instructive to consider interpretations based on the binary hypothesis since there are well-established binary accretion dynamics that might naturally explain such trends.

As the basis for an illustrative discussion on long-term trends in AGN LCs, we consider the binary hypothesis for PG 1553+113. In addition to near-orbital modulations in the jet luminosity, for certain binary parameters \citep[][]{dorazio_massratio, miranda_eccentricity, duffell_massratio, zrake_eccentricity}, a longer period is observed in circumbinary accretion simulations \citep[][]{macfadyen+2008}. This longer period is set by the orbital period associated with an overdensity (the ``lump'') in the circumbinary cavity wall \citep[see the description in, e.g.][]{miranda_eccentricity}. For year-like orbital periods appropriate for PG 1553+113, the lump period is measured to be $\approx 5-10$ orbital periods in simulations with baroclinic thermodynamics and radiative cooling \citep[e.g.][]{farris_trend_binary, WS_smbbh}. If the long-term MWL trends for PG 1553+113 are due to a lump, then visual inspection of the V-band currently indicates that the lump period is at least roughly eight orbits -- see Figure~\ref{fig:lc_pg_1553}. Whether the observed long-term trends are due to a lump implies different constraints on the binary parameters described in this section.

Firstly, we can derive a constraint on the binary eccentricity independently of the interpretation of the long-term MWL trends by showing that the binary is likely in the gravitational wave-driven (GW-driven) regime. A mass estimate of $(0.4-8)\times10^8 M_\odot$ for the central black hole in PG 1553+113 was made based on LC variability \citep{dhiman_variability}. Holding a binary SMBH's semi-major axis fixed, the binary is further from the GW-driven regime for lower total mass, higher accretion rate, lower eccentricity, and lower mass ratio \citep[][]{WS_smbbh}.
Thus, we can attempt to place the binary outside of the GW-driven regime by assuming its mass is the lower end of the above estimate $0.4\times10^8\, M_\odot$, it accretes at the Eddington limit (with a fiducial radiative efficiency of $\eta=0.1$), and it is circular eccentricity ($e=0$). We must also assume a fiducial accretion ``eigenvalue'' $l$ (we choose $l=1$) \citep[][]{paczynski_disk, popham_narayan_disk}, and we assume that the periodicity obtained in the LCs is roughly the redshifted orbital period $\approx 1.5 \times (1+z)$ yr for $z=0.433$. Doing so \citep[following][]{WS_smbbh}, we find the binary is in the GW-driven regime for all mass ratios $q\gtrsim 0.02$. \footnote{Note that if the binary instead had a super-Eddington accretion rate, which simulations suggest can be physically realized \citep[see, e.g.][]{Jiang+2019}, then the range of GW-driven mass ratios is narrower. For example, for $10\times$ Eddington accretion, the binary would be in the GW-driven regime for $q\gtrsim 0.25$.} Simulations show that mass ratios of accreting binaries are driven rapidly upward from such low values as $q=0.02$ \citep[e.g.][]{duffell_massratio}. Thus, the hypothetical binary in PG 1553+113 likely has $q>0.02$, and is, therefore, likely in the GW-driven regime. 

The fact that the hypothetical binary in PG 1553+113 is likely in the GW-driven regime, combined with simulation results, implies a constraint on the binary eccentricity. Simulations suggest an equilibrium value of binary eccentricity in the gas-driven regime of $e\approx 0.4$-$0.6$ \citep[][]{roedig_eccentricity1, roedig_eccentricity2, zrake_eccentricity, dorazio_eccentricity}, with the most recent results pointing to $e\approx0.4$. Since GW emission circularizes binaries, we expect the putative binary in PG~1553+113 to therefore have eccentricity $e\lesssim0.4$. This is our first constraint on the supposed binary. Note that simulations also predict that lump periodicity manifests when binaries have mass ratios $q\gtrsim 0.2$ \citep[e.g.][]{dorazio_massratio, duffell_massratio} and eccentricities $e\lesssim0.1$ \citep[e.g.][]{miranda_eccentricity, zrake_eccentricity}. 

We consider the following three scenarios for a hypothetical binary in PG 1553+113:
\begin{enumerate}
	\item The long-term MWL trends are due to a lump.
	\item The long-term MWL trends are not due to a lump because the lump is not present in the system.
	\item The long-term MWL trends are not due to a lump, even though the lump is present in the system.
\end{enumerate}
In the first scenario, a lump must exist; thus, the constraints on the binary from simulations are $q\gtrsim 0.2$ and $e\lesssim 0.1$. In the second scenario, the lump does not exist; thus, we either have a constraint on the mass ratio ($q\lesssim 0.2$) or a constraint on the eccentricity ($e\gtrsim 0.1$), or perhaps both. In either case, since the binary is likely in the GW-driven regime, the eccentricity is also constrained by $e\lesssim 0.4$. All of the constraints in this second scenario are consistent with the binary model proposed by \cite{cavaliere_binary_2017} ($q=0.1$, $e=0.2$), which was arrived at on the basis of different considerations. In this case, the long-term MWL trends are left unexplained and are presumably stochastic in the absence of another plausible mechanism. In the third scenario, lump periodicity does not transmit to jet variability in sufficient measure. 

Two mechanisms by which lump periodicity could imprint on blazar emission are via transmittance of lump periodicity to mass accretion rates \citep[a phenomenon which depends upon disk parameters; compare the simulations from, e.g.,][]{farris_trend_binary, WS_smbbh}, and periodic modulations in the supply of seed photons from the disk. In the accretion rate mechanism, one expects a systematic modulation of jet processes, manifesting in the SED via both the low-energy synchrotron component (e.g.~via modulation in the supply of electrons to the jet) and, consequently, the high-energy SSC component. In the seed photon mechanism, seed photons from the disk, modulated on the lump period, undergo inverse-Compton scattering in the jet(s) and/or disk corona \citep[``external Compton'' (EC, e.g.][]{band_synchrotron, levinson_jets}, thereby resulting in a modulation of the EC component of the SED on the lump period. The third scenario enumerated above would require that both of these imprint mechanisms are inefficient relative to other mechanisms of long-term variability. It is important to note that the SED of a BL Lac blazar like PG 1553+113 is often fit without an EC component \citep[see, e.g.,][]{aleksic_pg1553}, so if an EC component is non-negligible in the high-energy SED, then it must appear sufficiently degenerate with the SSC bump.

On the other hand, the first scenario enumerated above instead seems to require that both of these imprint mechanisms are efficient and comparably so. This is because the minimum of the long-term trend in optical, apparently occurring near 2015, does not appear to correspond to a minimum in $\gamma$-rays (see Figure~\ref{fig:lc_pg_1553}), although a longer temporal baseline is needed for higher confidence. Simulations have shown that there is a lag of order $\approx20-30\%$ of a lump period between lump modulations in thermal disk emission and accretion rates \citep[see Figure 5 in][]{farris_trend_binary}. Supposing that high-energy emission receives non-negligible contributions from both the SSC and EC components and that lower-energy emission does not receive a significant contribution from the EC component, one may therefore expect the lag between accretion and the external supply of seed photons could manifest as a multi-year shift of the minimum in the long-term MWL trend in $\gamma$-rays with respect to e.g.~optical \& UV.

Future theoretical work should address the efficiency of the mechanisms by which lump periodicity can imprint on blazar SEDs. If the long-term MWL trends in PG 1553+113 are due predominantly to lump periodicity, then we expect the recent upward trends reverse over the next few years, which is a strong motivation to continue monitoring PG~1553+113 across the entire electromagnetic spectrum.

\section{Summary} \label{sec:summary}
In this work, we have implemented a variability study of the MWL (radio, IR, optical, UV, and X ray) emission of the five blazars identified as having evidence of periodicity in \citet{penil_2022}. We find that two of them, PG 1553+113 and PKS 2155$-$304 show similar periodic behavior in their MWL emissions. The correlation $\gamma$ ray-optical does not indicate any lags with the limitation of 28 days due to the sampling of the \textit{Fermi}-LAT LCs. This result suggests a common spatial origin of both emissions. 
In the $\gamma$ ray-radio correlation, 0 and $\approx$200 days of delay have been observed in two blazars. No significant correlation above 3$\sigma$ is observed between $\gamma$-rays and X-rays. Regarding PG~1553+113, we made the first analysis of a long-term trend of increasing/decreasing flux in all bands. We explored an interpretation in terms of a hypothetical supermassive black hole binary central engine.

\section{Acknowledgements}
The \textit{Fermi} LAT Collaboration acknowledges generous ongoing support from a number of agencies and institutes that have supported both the development and the operation of the LAT as well as scientific data analysis. These include the National Aeronautics and Space Administration and the Department of Energy in the United States, the Commissariat \`a l'Energie Atomique and the Centre National de la Recherche Scientifique / Institut National de Physique Nucl\'eaire et de Physique des Particules in France, the Agenzia Spaziale Italiana and the Istituto Nazionale di Fisica Nucleare in Italy, the Ministry of Education, Culture, Sports, Science and Technology (MEXT), High Energy Accelerator Research Organization (KEK) and Japan Aerospace Exploration Agency (JAXA) in Japan, and the K.~A.~Wallenberg Foundation, the Swedish Research Council and the Swedish National Space Board in Sweden. Additional support for science analysis during the operations phase is gratefully acknowledged from the Istituto Nazionale di Astrofisica in Italy and the Centre National d'\'Etudes Spatiales in France. This work performed in part under DOE Contract DE-AC02-76SF00515.

P.P. and M.A. acknowledge funding under NASA contract 80NSSC20K1562. 

S.B. acknowledges financial support by the European Research Council for the ERC Starting grant MessMapp, under contract no. 949555.

A.D. is thankful for the support of the Ram{\'o}n y Cajal program from the Spanish MINECO, Proyecto PID2021-126536OA-I00 funded by MCIN / AEI / 10.13039/501100011033, and Proyecto PR44/21‐29915 funded by the Santander Bank and Universidad Complutense de Madrid.


L.M. acknowledges that support for this work was provided by NASA through
the NASA Hubble Fellowship grant \#HST-HF2-51486.001-A awarded by the
Space Telescope Science Institute, which is operated by the Association
of Universities for Research in Astronomy, Inc., for NASA, under
contract NAS5-26555.

J.O.S thanks the support from grant FPI-SO from the Spanish Ministry of Economy and Competitiveness (MINECO) (research project SEV-2015-0548-17-3 and predoctoral contract BES-2017-082171) and financial support from the Spanish Ministry of Science and Innovation (MICINN) through the Spanish State Research Agency, under Severo Ochoa Programme 2020-2023 (CEX2019-000920-S) and the project PID2019-107988GB-C22.
J.O.S. acknowledges financial support from the Severo Ochoa grant CEX2021-001131-S funded by MCIN/AEI/ 10.13039/501100011033. 

We also acknowledge support from NSF grant AST-1715661.

We also want to thank all the observatories from which we used data.  We thank the Las Cumbres Observatory and its staff for their continuing support of the ASAS-SN project. ASAS-SN is supported by the Gordon and Betty Moore Foundation through grant GBMF5490 to the Ohio State University, and NSF grants AST-1515927 and AST-1908570. Development of ASAS-SN has been supported by NSF grant AST-0908816, the Mt. Cuba Astronomical Foundation, the Center for Cosmology and AstroParticle Physics at the Ohio State University, the Chinese Academy of Sciences South America Center for Astronomy (CAS-SACA), the Villum Foundation, and George Skestos. The AAVSO database: Kafka, S., 2021, Observations from the AAVSO International Database, \url{https://www.aavso.org}. The CSS survey is funded by the National Aeronautics and Space Administration under Grant No. NNG05GF22G was issued through the Science Mission Directorate Near-Earth Objects Observations Program. The Catalina Real-Time Transient Survey is supported by the U.S.~National Science Foundation under grants AST-0909182 and AST-1313422. This paper has made use of up-to-date SMARTS optical/near-infrared light curves that are available at \url{www.astro.yale.edu/smarts/glast/home.php}. Data from the Steward Observatory spectropolarimetric monitoring project were used. This program is supported by Fermi Guest Investigator grants NNX08AW56G, NNX09AU10G, NNX12AO93G, and NNX15AU81G. This research has made use of data from the OVRO 40-m monitoring program \citep{ovro_monitoring}, supported by private funding from the California Institute of Technology and the Max Planck Institute for Radio Astronomy, and by NASA grants NNX08AW31G, NNX11A043G, and NNX14AQ89G and NSF grants AST-0808050 and AST-1109911. We also acknowledge the use of public data from the {\it Swift} data archive.

\section{Data Availability}

All the data used in this work are publicly available or available on request to the responsible for the corresponding observatory/facility. All the links to the databases, online repositories, and/or contact information are provided in the footnotes in Section \ref{sec:wave_data}.



\bibliographystyle{mnras}
\bibliography{literatura} 

\clearpage

\begin{table*}
\centering
\caption{List of periods and their associated uncertainties (superscript),  local significances (left subscript), and global significances (right subscript) of PG 1553+113 and PKS 2155$-$304. The symbol "$\star$" is utilized to represent two periods that have been derived from the same energy band and exhibit similar significance. These periods are organized based on their peak amplitudes. The symbol $\dagger$ is used to denote the periods where the PDM results exhibit the harmonic effect as described in P20. The symbol $\ddagger$ is used to indicate periods that are consistently presented in the WWZ for all the LC. The MCMC sine fitting results provide information solely about the inferred period and its associated uncertainties. The period values are expressed in years.}
\label{tab:methods_results}
{%
\begin{tabular}{c|cccccccccc}
\hline
\hline
Association & Wavelength & LSP + & GLSP & LSP + & PDM & WWZ & DFT-Welch & MCMC Sine \\
 Name &  & Power-Law &  & Simulated LC &  &  &  & Fitting \\
 \hline
        PG 1553+113 &
	\makecell{\makecell{{\it Swift}/XRT$\star$ \\  \\ } \\[3pt] UVOT ({\it uvw2}) \\[3pt] UVOT ({\it uvm2}) \\[3pt] UVOT ({\it uvw1}) \\[3pt] V-BAND \\[3pt] R-BAND \\[3pt] OVRO} &
	\makecell{\makecell{$1.5^{\pm0.5}_{3.1\sigma/0.0\sigma}$ \\[2pt] $2.3^{\pm0.7}_{2.8\sigma/0.0\sigma}$} \\[2pt] $2.2^{\pm0.2}_{3.0\sigma/0.0\sigma}$ \\[3pt] $2.2^{\pm0.2}_{3.0\sigma/0.0\sigma}$ \\[3pt] $2.2^{\pm0.2}_{3.0\sigma/0.0\sigma}$ \\[3pt] $2.3^{\pm0.2}_{5.3\sigma/3.1\sigma}$ \\[3pt] $2.3^{\pm0.2}_{5.2\sigma/3.0\sigma}$ \\[3pt] $2.2^{\pm0.3}_{5.1\sigma/2.9\sigma}$} &
	\makecell{\makecell{$1.5^{\pm0.1}_{3.6\sigma/0.0\sigma}$ \\ \\} \\[3pt] $1.4^{\pm0.1}_{1.8\sigma/0.0\sigma}$ \\[3pt] $1.4^{\pm0.2}_{1.9\sigma/0.0\sigma}$ \\[3pt] $4.9^{\pm0.5}_{2.0\sigma/0.0\sigma}$ \\[3pt] $2.3^{\pm0.2}_{4.0\sigma/0.6\sigma}$ \\[3pt] $2.1^{\pm0.3}_{4.1\sigma/0.8\sigma}$ \\[3pt] $2.2^{\pm0.2}_{5.2\sigma/3.0\sigma}$} &
	\makecell{\makecell{$1.5^{\pm0.2}_{1.9\sigma/0.0\sigma}$ \\[2pt] $2.3^{\pm0.2}_{2.5\sigma/0.0\sigma}$} \\[3pt] $2.2^{\pm0.2}_{4.5\sigma/1.8\sigma}$ \\[3pt] $2.2^{\pm0.2}_{4\sigma/0.6\sigma}$ \\[3pt] $2.2^{\pm0.2}_{5.2\sigma/3.0\sigma}$ \\[3pt] $2.3^{\pm0.2}_{5.0\sigma/2.9\sigma}$ \\[3pt] $2.1^{\pm0.2}_{4.2\sigma/1.1\sigma}$ \\[3pt] $2.2^{\pm0.2}_{5.1\sigma/2.9\sigma}$} &
	\makecell{\makecell{$\dagger$$3.0^{\pm0.2}_{2.9\sigma/0.0\sigma}$ \\ [5pt] } \\ [3pt] $\dagger$$4.4^{\pm0.3}_{3.4\sigma/0.0\sigma}$ \\[3pt] $\dagger$$4.4^{\pm0.3}_{2.9\sigma/0.0\sigma}$ \\[3pt] $\dagger$$4.3^{\pm0.3}_{2.9\sigma/0.0\sigma}$ \\[3pt] $2.2^{\pm0.5}_{3.1\sigma/0.0\sigma}$ \\[3pt] $\dagger$$4.4^{\pm0.4}_{3.1\sigma/0.0\sigma}$ \\[3pt] $2.4^{\pm0.5}_{3.5\sigma/0.0\sigma}$} &
	\makecell{\makecell{$5.3^{\pm1.2}_{0.9\sigma/0.0\sigma}$ \\ \\} \\ $2.1^{\pm0.1}_{1.3\sigma/0.0\sigma}$ \\[3pt] $2.0^{\pm0.1}_{1.6\sigma/0.0\sigma}$ \\[3pt] $2.0^{\pm0.2}_{1.3\sigma/0.0\sigma}$ \\[3pt] $2.2^{\pm0.3}_{5.0\sigma/2.8\sigma}$ \\[3pt] $\ddagger$$2.2^{\pm0.6}_{5.0\sigma/2.8\sigma}$ \\[3pt] $\ddagger$$2.2^{\pm0.2}_{5.1\sigma/3.0\sigma}$} &
	\makecell{\makecell{$1.5^{\pm0.1}_{2.0\sigma/0.0\sigma}$ \\ \\} \\[3pt] $2.1^{\pm0.3}_{2.4\sigma/0.0\sigma}$ \\[3pt] $2.1^{\pm0.3}_{2.0\sigma/0.0\sigma}$ \\[3pt] $2.2^{\pm0.3}_{2.0\sigma/0.0\sigma}$ \\[3pt] $2.2^{\pm0.2}_{4.0\sigma/0.6\sigma}$ \\[3pt] $2.1^{\pm0.3}_{4.1\sigma/0.8\sigma}$ \\[3pt] $2.2^{\pm0.3}_{4.9\sigma/2.5\sigma}$} &
	\makecell{\makecell{$1.5^{+0.8}_{-0.1}$ \\ \\} \\ 2.1$\pm$0.1 \\ 2.1$\pm$0.1 \\[3pt] 2.1$\pm$0.1 \\[3pt] 2.3$\pm$0.1 \\[3pt] 2.1$\pm$0.1 \\[3pt] 2.3$\pm$0.3} \\[3pt]
        \hline
        PKS 2155$-$304 &
	\makecell{{\it Swift}/XRT \\[3pt] UVOT ({\it uvw2}) \\[3pt] UVOT ({\it uvm2}) \\[3pt] UVOT ({\it uvw1}) \\[3pt] SMARTS-B \\[3pt] V-BAND \\[3pt] R-BAND \\[3pt] SMARTS-J \\[3pt] SMARTS-K} &
	\makecell{$1.7^{\pm0.2}_{2.8\sigma/0.0\sigma}$ \\[3pt] $1.8^{\pm0.1}_{3.4\sigma/0.0\sigma}$ \\[3pt] $1.8^{\pm0.1}_{2.8\sigma/0.0\sigma}$ \\[3pt] $1.8^{\pm0.1}_{2.9\sigma/0.0\sigma}$ \\[3pt] $1.6^{\pm0.1}_{3.6\sigma/0.0\sigma}$ \\[3pt] $1.8^{\pm0.1}_{4.4\sigma/1.5\sigma}$ \\[3pt] $1.6^{\pm0.2}_{3.1\sigma/0.0\sigma}$ \\[3pt] $1.7^{\pm0.1}_{3.0\sigma/0.0\sigma}$ \\[3pt] $1.7^{\pm0.3}_{2.1\sigma/0.0\sigma}$} &
	\makecell{$1.6^{\pm0.2}_{2.7\sigma/0.0\sigma}$ \\[3pt] $1.8^{\pm0.1}_{1.8\sigma/0.0\sigma}$ \\[3pt] $1.2^{\pm0.1}_{1.4\sigma/0.0\sigma}$ \\[3pt] $1.4^{\pm0.1}_{1.1\sigma/0.0\sigma}$ \\[3pt] $3.6^{\pm1.2}_{1.6\sigma/0.0\sigma}$ \\[3pt] $1.7^{\pm0.1}_{2.5\sigma/0.0\sigma}$ \\[3pt] $1.7^{\pm0.1}_{2.3\sigma/0.0\sigma}$ \\[3pt] -- \\[3pt] $1.7^{\pm0.4}_{1.2\sigma/0.0\sigma}$} &
	\makecell{$1.8^{\pm0.2}_{3.2\sigma/0.0\sigma}$ \\[3pt] $1.9^{\pm0.1}_{3.0\sigma/0.0\sigma}$ \\[3pt] $1.8^{\pm0.1}_{2.0\sigma/0.0\sigma}$ \\[3pt] $1.8^{\pm0.2}_{2.1\sigma/0.0\sigma}$ \\[3pt] $1.7^{\pm0.1}_{3.1\sigma/0.0\sigma}$ \\[3pt] $1.8^{\pm0.1}_{3.6\sigma/0.0\sigma}$ \\[3pt] $1.8^{\pm0.1}_{3.0\sigma/0.0\sigma}$ \\[3pt] $1.8^{\pm0.1}_{3.6\sigma/0.0\sigma}$ \\[3pt] $1.8^{\pm0.3}_{2.0\sigma/0.0\sigma}$} &
	\makecell{$3.9^{\pm0.2}_{1.6\sigma/0.0\sigma}$ \\[3pt] $1.9^{\pm0.1}_{2.5\sigma/0.0\sigma}$ \\[3pt] $1.9^{\pm0.1}_{2.2\sigma/0.0\sigma}$ \\[3pt] $1.9^{\pm0.1}_{2.1\sigma/0.0\sigma}$ \\[3pt] $3.3^{\pm0.1}_{2.6\sigma/0.0\sigma}$ \\[3pt] $\dagger$$3.2^{\pm0.2}_{2.1\sigma/0.0\sigma}$ \\[3pt] $\dagger$$3.3^{\pm0.2}_{3.1\sigma/0.0\sigma}$ \\[3pt] $\dagger$$3.3^{\pm0.1}_{2.4\sigma/0.0\sigma}$ \\[3pt] $1.7^{\pm0.2}_{2.8\sigma/0.0\sigma}$} &
	\makecell{$1.6^{\pm0.1}_{2.1\sigma/0.0\sigma}$ \\[3pt] $1.9^{\pm0.5}_{1.6\sigma/0.0\sigma}$ \\[3pt] $1.9^{\pm0.5}_{1.4\sigma/0.0\sigma}$ \\[3pt] $1.9^{\pm0.5}_{1.4\sigma/0.0\sigma}$ \\[3pt] $\ddagger$$1.7^{\pm0.6}_{2.4\sigma/0.0\sigma}$ \\[3pt] $1.7^{\pm0.6}_{2.4\sigma/0.0\sigma}$ \\[3pt] $\ddagger$$1.8^{\pm0.7}_{2.3\sigma/0.0\sigma}$ \\[3pt] $\ddagger$$1.8^{\pm0.8}_{2.0\sigma/0.0\sigma}$ \\[3pt] $1.7^{\pm0.6}_{1.6\sigma/0.0\sigma}$} &
	\makecell{$1.6^{\pm0.2}_{2.3\sigma/0.0\sigma}$ \\[3pt] $1.5^{\pm0.1}_{1.0\sigma/0.0\sigma}$ \\[3pt] $1.6^{\pm0.1}_{1.2\sigma/0.0\sigma}$ \\[3pt] $1.6^{\pm0.1}_{1.2\sigma/0.0\sigma}$ \\[3pt] $1.6^{\pm0.1}_{4.0\sigma/0.6\sigma}$ \\[3pt] $1.6^{\pm0.2}_{2.4\sigma/0.0\sigma}$ \\[3pt] $1.6^{\pm0.1}_{2.4\sigma/0.0\sigma}$ \\[3pt] $1.9^{\pm0.2}_{2.0\sigma/0.0\sigma}$ \\[3pt] $1.8^{\pm0.3}_{1.7\sigma/0.0\sigma}$} &
	\makecell{$1.8^{+0.9}_{-0.1}$ \\[3pt] $1.8^{+0.1}_{-0.5}$ \\[3pt] 2.0$\pm$0.5 \\[3pt] 2.1$\pm$0.5 \\[3pt] $1.7^{+0.5}_{-0.1}$ \\[3pt] $1.8^{+0.9}_{-0.5}$ \\[3pt] $1.8^{+0.6}_{-0.1}$ \\[3pt] 1.8$\pm$0.5 \\[3pt] 1.7$\pm$0.6} \\
 \hline
 \hline
\end{tabular}%
}
\end{table*}
\begin{table*}
\centering
\caption{Results of the \textit{z}-DCF cross-correlation analysis for PG 1553+113 and PKS 2155$-$304. The table also shows the period inferred from the cross-correlation with the $\gamma$-ray LC and the auto-correlation. These periods are associated with the significance levels of the oscillations (high-level, low-level, see Figure \ref{fig:autocorrelation}). We report one significance when the high and low significance levels are the same. The correlation with the $\gamma$ rays is limited by the 28 days sampling of the \textit{Fermi}-LAT LCs.}
\label{tab:cross_correlation}
{%
\begin{tabular}{c|ccccccc}
\hline
\hline
Association & Wavelength & Correlation & Cross-correlation & Auto-correlation  \\
 Name &  & Lag {[}days{]} & Period {[}years{]} & Period {[}years{]} \\
 \hline
 	PG 1553+113 &
	\makecell{\textit{Fermi}-LAT \\ {\it Swift}/XRT \\ UVOT ({\it uvw2}) \\ UVOT ({\it uvm2}) \\ UVOT ({\it uvw1})  \\ V-BAND \\ R-BAND \\ OVRO} &
	\makecell{-- \\ $1.5^{\pm0.1}_{\approx3.0\sigma}$ \\[2pt] $1.5^{\pm0.1}_{>4.0\sigma}$ \\[2pt] $1.5^{\pm0.1}_{>4.0\sigma}$ \\[2pt] $1.5^{\pm0.1}_{\approx4.0\sigma}$ \\[2pt] $15.8^{\pm0.8}_{>2.0\sigma}$ \\[2pt] $0.5^{\pm0.1}_{>2.0\sigma}$ \\[2pt] $238.1^{\pm9.1}_{>2.0\sigma}$} &
	\makecell{-- \\ $2.2^{\pm0.2}_{(3-2)\sigma}$ \\[2pt]  $2.1^{\pm0.1}_{(3-1)\sigma}$ \\[2pt] $2.1^{\pm0.1}_{(3-1)\sigma}$ \\[2pt] $2.1^{\pm0.1}_{(3-1)\sigma}$ \\[2pt] $2.3^{\pm0.1}_{2\sigma}$ \\[2pt] $2.0^{\pm0.1}_{2\sigma}$ \\[2pt] $2.1^{\pm0.2}_{2\sigma}$} &
	\makecell{$2.2^{\pm0.1}_{(3-2)\sigma}$ \\ $1.4^{\pm0.2}_{(2-1)\sigma}$ \\ $2.0^{\pm0.1}_{(2-1)\sigma}$ \\ $2.0^{\pm0.2}_{(2-1)\sigma}$ \\ $2.0^{\pm0.2}_{(2-1)\sigma}$ \\  $2.2^{\pm0.1}_{2\sigma}$ \\ $2.1^{\pm0.1}_{(3-2)\sigma}$ \\ $2.2^{\pm0.1}_{2\sigma}$} 
	\\
	\hline
	PKS 2155$-$304 &
	\makecell{\textit{Fermi}-LAT \\ {\it Swift}/XRT \\ UVOT ({\it uvw2}) \\ UVOT ({\it uvm2}) \\ UVOT ({\it uvw1}) \\ SMARTS-B \\ V-BAND \\ R-BAND \\ SMARTS-J \\ SMARTS-K} &
	\makecell{--\\ $10.5^{\pm0.1}_{\approx3.1\sigma}$ \\[2pt] $-10.8^{\pm0.3}_{5.0\sigma}$ \\[2pt] $-10.8^{\pm0.3}_{5.3\sigma}$ \\[2pt] $-10.6^{\pm0.3}_{5.2\sigma}$ \\[2pt]  $15.6^{\pm0.4}_{5.0\sigma}$ \\[2pt] $6.9^{\pm0.1}_{5.1\sigma}$ \\[2pt] $4.8^{\pm0.1}_{5.1\sigma}$ \\[2pt] $15.6^{\pm0.4}_{3.9\sigma}$ \\[2pt] $-4.6^{\pm9.8}_{3.7\sigma}$} &
	\makecell{--\\ $1.7^{\pm0.1}_{(2-1)\sigma}$ \\[2pt] $1.7^{\pm0.1}_{(3-1)\sigma}$ \\[2pt] $1.7^{\pm0.1}_{(3-1)\sigma}$ \\[2pt] $1.7^{\pm0.1}_{(3-1)\sigma}$ \\[2pt] $1.6^{\pm0.2}_{(1-3)\sigma}$ \\[2pt] $1.6^{\pm0.2}_{(1-3)\sigma}$ \\[2pt] $1.6^{\pm0.2}_{(1-3)\sigma}$ \\[2pt] $1.8^{\pm0.1}_{(1-3)\sigma}$ \\ --} &
	\makecell{$1.7^{\pm0.2}_{(1-2)\sigma}$ \\[2pt] $1.9^{\pm0.2}_{1\sigma}$ \\[2pt] $1.8^{\pm0.3}_{(2-1)\sigma}$ \\[2pt] -- \\ $1.7^{\pm0.4}_{(3-1)\sigma}$ \\[2pt] $1.5^{\pm0.2}_{(2-1)\sigma}$ \\[2pt] $1.7^{\pm0.2}_{(2-1)\sigma}$ \\[2pt] $1.5^{\pm0.1}_{1\sigma}$ \\ -- \\ --} \\
\hline
\hline
\end{tabular}%
}
\end{table*}



\clearpage

\appendix
\renewcommand{\thesubsection}{\Alph{subsection}}
\renewcommand{\thefigure}{A\arabic{figure}}
\renewcommand{\thetable}{A\arabic{table}}
\section*{Appendix}\label{sec:appendix}
\subsection{Periodicity and correlation results}
\subsubsection{Periodicity} \label{sec:periodicity_appendix}

The results of the periodicity analysis are shown in Table \ref{tab:methods_results_appendix}.
PKS 0454$-$234 has no significant evidence of periodicity inferred from the analysis in the available bands. However, the optical data have insufficient temporal coverage ($\approx$7.0 yr, see Figure \ref{fig:mwl_pks0454}) to reliably rule out the period observed in $\gamma$-rays ($\approx$3.6 yr). 

For S5~0716$+$71 (see Figure \ref{fig:mwl_s50716}), we find a period compatible with that observed in $\gamma$-rays and in the V-band, $\approx$2.7 yr \citep[similar to][]{bhatta_s50714_optical}. A period of $\approx$1 yr is also inferred by some of the employed methods. This is compatible with the results from P22, where both periods were obtained in $\gamma$-rays. However, this V-band period of $\approx$2.7 yr is not significant ($<$3$\sigma$ of local significance). \citet{sandrinelli_S5_0716_71} report no significant evidence of periodicity in the R-band. We do not observe significant periodicity in X-ray and radio bands. 

We do not find any evidence of periodicity for OJ 014 in the radio band. The periodicity analysis for the optical band was not performed due to insufficient temporal coverage of the data (Figure \ref{fig:mwl_oj014}).

\subsubsection{Correlation} \label{sec:correlations_results_appendix}
Correlation results are shown in Table \ref{tab:cross_correlation_appendix}, including the periods inferred with the \textit{z}-DCF.

For PKS 0454$-$234, \citet{cor_kait_bigsample} claim a delay between $\gamma$-ray and the optical bands of $\approx$150 days; instead, we find a zero lag with local significance of $>$4$\sigma$. No period is inferred from the autocorrelation, in agreement with the results in $\S$\ref{sec:periodicity_appendix}.

Regarding S5 0716+71, we find a lag of $\approx$0 days in the X-ray, optical, and radio LCs ($\geq$2$\sigma$). Additionally, the cross-correlation analysis suggests a $\approx$2.7 yr period for both optical and radio bands, compatible with the one from P22, but not significant. The auto-correlation also suggests a compatible period for the V-band, but not at a significant level. 

OJ 014 presents no time lag between its $\gamma$-ray and optical emission. For radio, the lag is $-$40 days ($>$4$\sigma$). Both correlations show a compatible period with $\gamma$-rays but it is not significant.

In some cases, the results of Table \ref{tab:methods_results_appendix} and the autocorrelations are significantly different (e.g., S5 0716+71). There are several factors that contribute to the divergent results. The shorter time coverage and uneven sampling of the MWL data can lead to much larger errors in the derived periods than those derived from the Fermi-LAT LCs, including resulting in the absence of a period (note that all of the autocorrelations for $\gamma$-rays are compatible with those reported in P22). Additionally, the methods handle LC gaps differently (see P20).  Finally, each method is impacted differently by the choice of binning in each analysis \citep[periodicity and autocorrelation][]{otero_2023}.

\subsection{Variability Study}\label{sec:variability}
LC variability is studied using the fractional variability ($F_{var}$), the structure function (SF), as well as the PSD and PDF. $F_{var}$ and the SF are respectively used to quantify the variability \citep{vaughan_fractional_variability} and measure the characteristic variability timescales. $F_{\text{var}}$ is affected by the time coverage, sampling, and binning of the data \citep[see][]{vaughan_fractional_variability, schleicher_variability}. Consequently, employing a similar time window and binning to analyze MWL data. Additionally, using nonsimultaneous data can also lead to inconsistencies in  $F_{\text{var}}$ \citep[][]{schleicher_variability}.

Finally, We also analyze the polarization of blazar emission using polarimetric data from the Steward Observatory.

\subsubsection{Fractional Variability}
We have evaluated the amount of variation displayed by our blazar sample through the evaluation of $F_{\text{var}}$. The results are presented in Table \ref{tab:fractional}. The $F_{\text{var}}$ values, typically $>$0.25, prove the variable nature of the sources studied here. We also observe that the most variable source of the sample (that with the highest $F_{\text{var}}$ in the different bands) is the only FSRQ included in this study, PKS 0454$-$234. On the other hand, BL Lac objects tend to have smaller values of $F_{\text{var}}$ \citep[e.g.,][]{bhatta_s5_0716}. In fact, for the four BL Lac objects included here, the two showing the synchrotron peak at lowest frequencies (OJ 014 and S5 0716+714) are those with the highest $F_{\text{var}}$ among the BL Lacs, while PG 1553+113 and PKS 2155$-$304, with higher synchrotron peak frequencies, are less variable in our study. Despite the small number of sources studied here, this trend of increasing $F_{\text{var}}$ with the decreasing value of $\nu_{sync}$ is in line with the results reported in other studies \citep[e.g.,][]{bhatta_s5_0716}. Hence, FSRQs and low-synchrotron-peak BL Lacs are typically more variable than high-synchrotron-peak BL Lacs.

\begin{figure*}[ht!]
\includegraphics[scale=0.22]{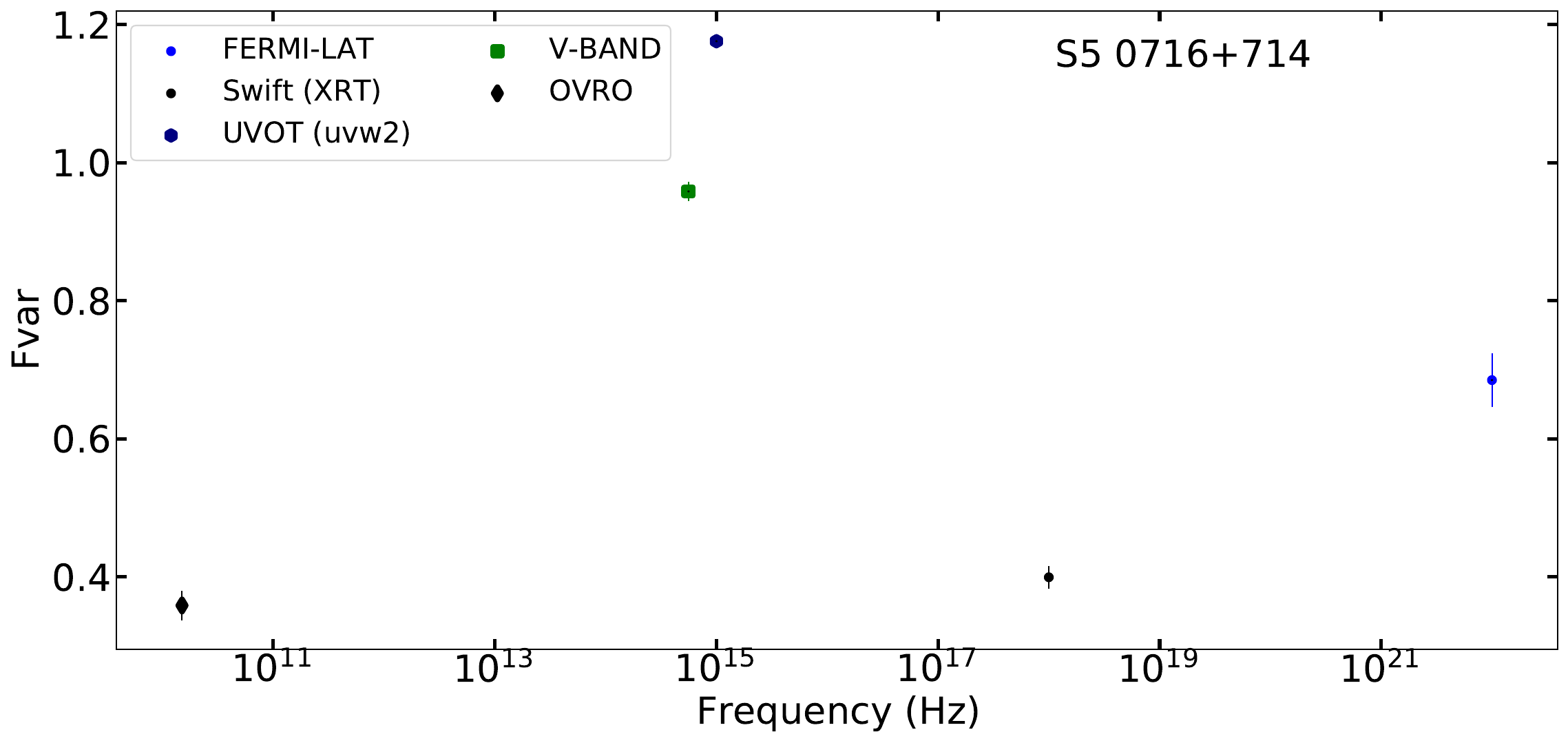}
\includegraphics[scale=0.22]{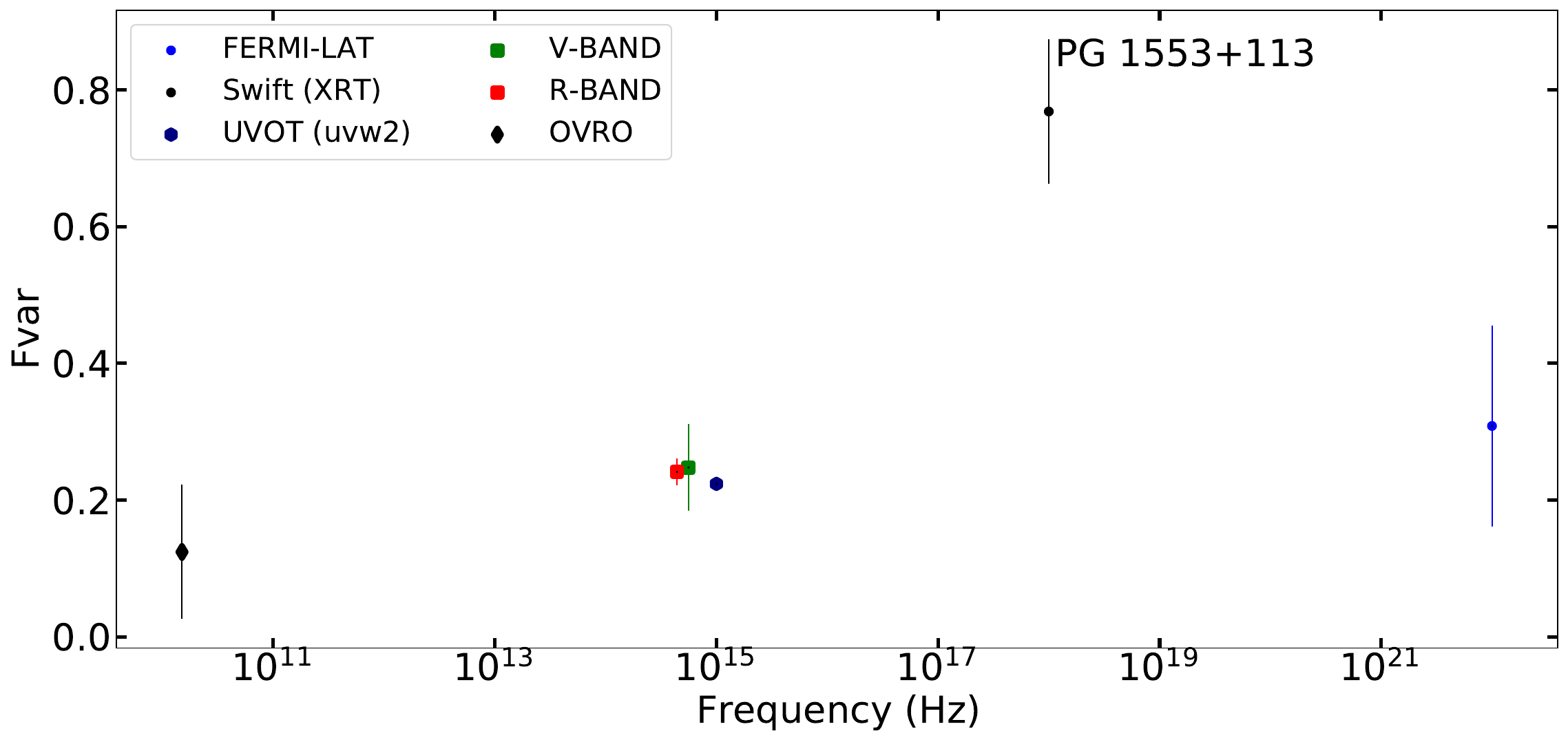}
\caption{Fractional variability for the MWL simultaneous data (taken within the same 24-hour period) for S5~0716+714 (\textit{left}), and PG~1553+113 (\textit{right}).}
\label{fig:fvar}
\end{figure*}

We note that the calculation and comparison of the MWL $F_{\text{var}}$ might be highly affected by the non-simultaneity of the data between the different bands \citep[][]{schleicher_variability}. Therefore, as a cross-check, we have calculated $F_{\text{var}}$ using only the simultaneous data in all the bands.  $F_{\text{var}}$ derived for simultaneous data (i.e., taken within the same 24-hour period) yields values with no significant differences in comparison to $F_{\text{var}}$ obtained with all the data of the LCs (see figures of $\S$\ref{sec:plots}), except for the case of the X-ray $F_{\text{var}}$ values of PG~1553+113 and S5~0716+714, which $F_{\text{var}}$ of the simultaneous data are $\approx$15\% higher than $F_{\text{var}}$ obtained with all the X-ray data (see Figure \ref{fig:fvar}). Nevertheless, the evolution with the frequency of $F_{\text{var}}$ remains unchanged with both approaches.

By studying the evolution with the frequency of $F_{\text{var}}$ in the different bands, we can also retrieve information on the processes and particle populations dominating the variability. However, for the five sources analyzed here, the MWL coverage is not always optimal for evaluating the evolution of $F_{\text{var}}$ with the frequency. Nevertheless, we observe the minimum variability for PG~1553+113 at radio wavelengths (see Figure \ref{fig:fvar}), increasing up to its maximum in the X-ray band. Then, $F_{\text{var}}$ decreases in the HE $\gamma$-ray band. We do not have data in the very-high-energy (VHE, E$>$100~GeV) $\gamma$-ray band. We note that \cite{aleksic_pg1553_fvar} report for VHE an $F_{\text{var}}$ close to that in the X-ray band. This could point towards a double-peaked shape, observed in the past for several sources \citep{aleksic_variability}. Since this source is a high-synchrotron-peak BL Lac object with its X-ray emission corresponding to the end of the synchrotron bump, a higher variability in this regime and in the VHE $\gamma$-ray band could indicate a higher variability of the high-energy electron population responsible for the emission. In comparison, we can see the case of S5~0716+714 (see Figure \ref{fig:fvar}), where the maximum, $\approx$1.2, is found at UV wavelengths, in comparison with its minimum in the X-ray band. In this case, the source is an intermediate-synchrotron-peak BL Lac, and its X-ray emission is found in the transition between the low- and high-energy SED peaks. Hence, depending on $F_{\text{var}}$ structure and SED characteristics, we can observe differences between sources that may reveal information regarding the particle populations responsible for the emission and variability observed.

\subsubsection{Characteristic timescales}

\begin{figure}	        
    \includegraphics[width=\columnwidth]{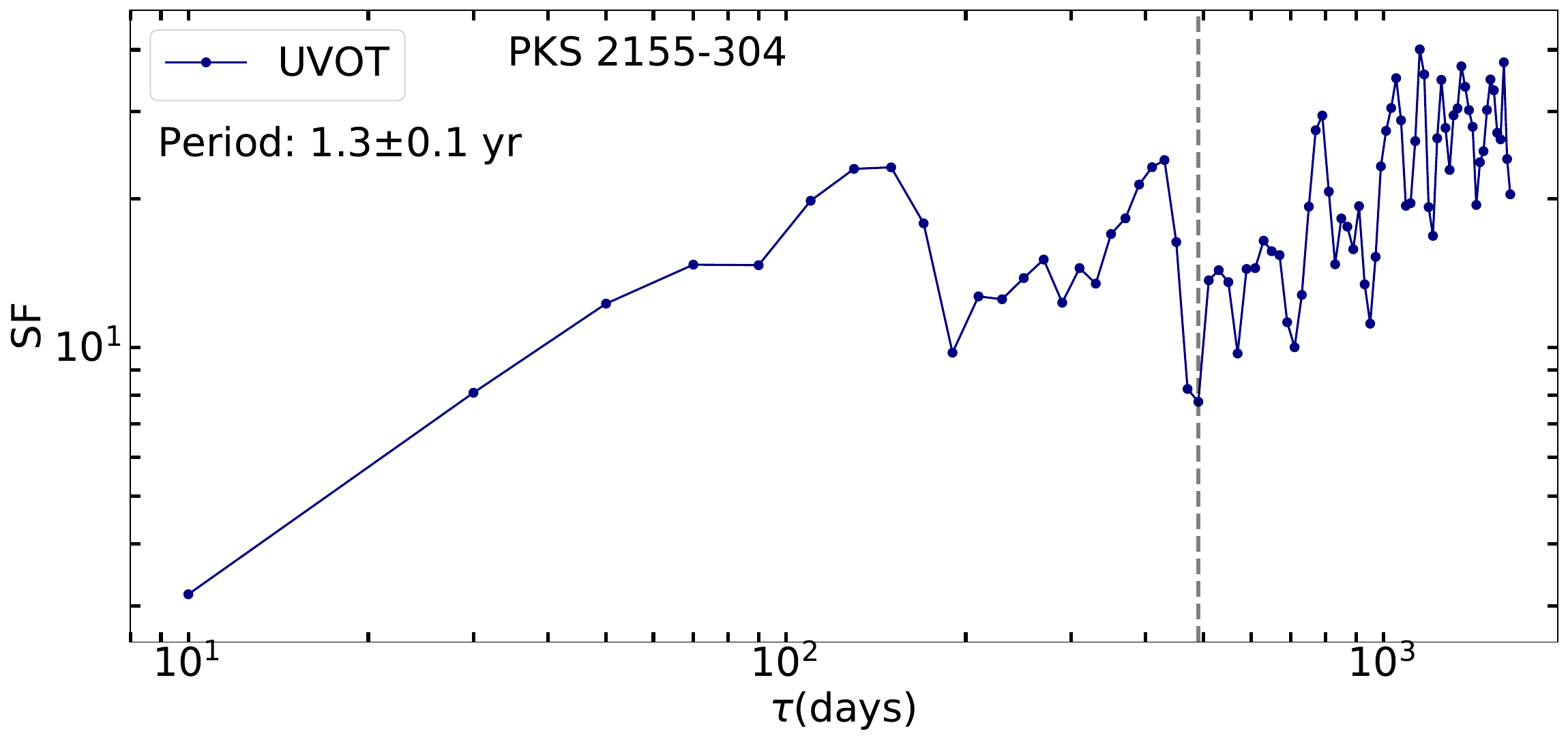}
    \caption{Structure function of the UVOT LC of PKS~2155$-$304. The vertical line shows the dip that denotes the presence of a period in the signal.}
    \label{fig:sf}
\end{figure}

The results of Table \ref{tab:fractional} reveal different variability timescales, ranging from $\approx$100 days to $\approx$4 yr. The emission is most stable in the radio. The most variable is in the optical, as inferred by the set of timescales derived from the structure functions (SF, see Figure \ref{fig:sf}).

The average slope of the SFs analyzed in this work is 0.5$\pm$0.1. This result is compatible with a slope of 0.44$\pm$0.03 associated with a model based on instabilities in the accretion disk, as derived by \citet{hawkins_sf}. Regarding the periodicity, most of the periods inferred with the SF (see Table \ref{tab:fractional}) are compatible with the ones shown in Table \ref{tab:methods_results} and Table \ref{tab:methods_results_appendix}.  

\subsubsection{Power Spectrum Index Estimation}
The results of the power spectrum index estimation are shown in Table \ref{tab:slopes}. The slopes obtained for the optical and radio bands range between 1.2 and 1.5, compatible with slopes of the $\gamma$-ray emission presented in P22. This index range can be associated with an admixture of flickering (pink) noise, consisting of short-scale variability and red noise associated with long-term variability. According to \citet{abdo_variability}, these oscillations are associated with instability and turbulence in the accretion flow through the disk or the jet.

\subsubsection{Polarized Light}
\begin{figure}	        
    \includegraphics[width=\columnwidth]{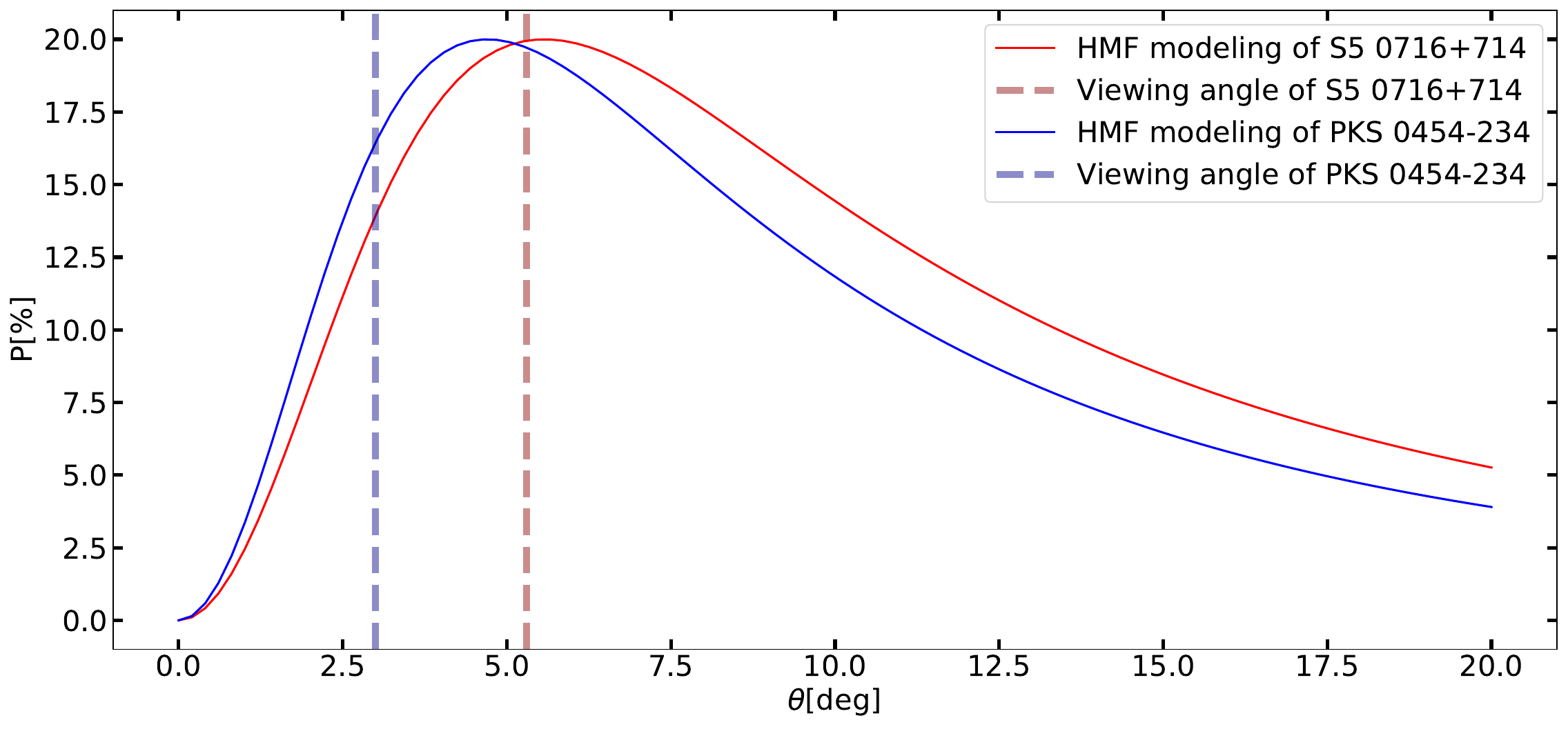}
    \caption{Helical magnetic field (HMF) polarization models for PKS~0454$-$234 and S5~0716+714 \citep[][]{raiteri_polarization_model}. The dashed lines denote the viewing angle $\rm \theta_{obs}$.}
    \label{fig:polarisation}
\end{figure}

Blazars are characterized by high levels of radio and optical polarization. This polarized emission is also extremely variable, and it depends on the structure of the magnetic field of the emitting region. Thus, it can provide valuable information about the origin of the emission in blazars. \citet{raiteri_polarization_model} interpret the long-term flux variability of the polarized optical emission according to two different models: helical magnetic fields and transverse shock waves. Both models predict two different behaviors of the polarization degree w.r.t. the viewing angle $\rm \theta$ (see Figure \ref{fig:polarisation}). In both models, when $\rm \theta<\theta_{max}$ (with $\rm \theta_{max} \approx 1/\Gamma$ the angle of maximum polarization and $\rm \Gamma$ the bulk Lorentz factor of the jet), the polarization degree increases with the viewing angle of the jet. The opposite is expected when $\rm \theta>\theta_{max}$ (distinguishing two regions, see Figure \ref{fig:polarisation}). Additionally, the observed flux increases for decreasing the viewing angle \citep{raiteri_polarization_model}. Therefore, in the first (second) region, the polarization degree is anti-correlated (correlated) with the observed flux. Consequently, the correlation between the polarized and total emission can constrain the scenarios responsible for the variability. Specifically, inferring a correlation for viewing angles $\rm <\theta_{max}$ between the polarized and total flux could indicate that the variability is not associated with internal-jet processes but produced by external phenomena affecting the jet \citep[e.g., the accretion disc, or the single/binary black hole system, ][]{raiteri_polarization_model}. 
In Figure \ref{fig:polarisation}, the polarization degree as a function of the viewing angle is represented for the helical magnetic field model \citep[the transverse shock model has a similar behavior, see Figure 17 in][]{raiteri_polarization_model}. We use $\Gamma=12.2$ and $\rm \theta_{obs}=3.0^{\circ}$ for PKS 0454$-$234 \citep{ghisellini_pks0454_polarization}, and $\Gamma=10.3$ and $\rm \theta_{obs}=5.2^{\circ}$ for S5~0716+714 \citep{hovatta_s50716_polarization}. Observing the plot for both blazars, polarization increases as the viewing angle increases from $\rm \theta$=0$^{\circ}$ to $\rm \theta_{max}$, and then slowly decreases for higher viewing angles.

In this context, we evaluate the correlation of V-band emission of PKS 0454$-$234 and S5 0716+714 between the polarized and total flux (since only these sources have available data in the Steward database). For PKS~0454$-$234, both emissions are correlated, with a time lag of $4.9\pm11.9$ days  ($>$2$\sigma$, significance before applying any trial correction, as in the other significance reported in this section). A similar result is observed for S5 0716+714, where a correlation is obtained, with a time lag of $-0.1{\pm8.5}$ days ($>$3$\sigma$). Consequently, the variability of both objects can not be explained by internal phenomena associated with the jet, according to the models in \citet{raiteri_polarization_model}. 

\begin{table*}
\centering
\caption{List of periods and their associated uncertainties (superscript), local significances (left subscript), and global significances (right subscript) of PKS 0454$-$234, S5 0716+714, OJ 014. The symbol "$\star$" is utilized to represent two periods that have been derived from the same energy band and exhibit similar significance. These periods are organized based on their peak amplitudes. The symbol $\dagger$ is used to denote the periods where the PDM results exhibit the harmonic effect as described in P20. The symbol $\ddagger$ is used to indicate periods that are consistently presented in the WWZ for all the LC. The MCMC sine fitting results provide information solely about the inferred period and its associated uncertainties. The period values are expressed in years.}
\label{tab:methods_results_appendix}
{%
\begin{tabular}{c|cccccccccc}
\hline
\hline
Association & Wavelength & LSP + & GLSP & LSP + & PDM & WWZ & DFT-Welch & MCMC Sine \\
 Name &  & Power-Law &  & Simulated LC &  &  &  & Fitting \\
 \hline
PKS 0454$-$234 & 
	\makecell{V-BAND  \\[2pt] STEWARD-V \\[2pt] STEWARD-R \\[2pt] R-BAND$\star$ } &
	\makecell{$1.8^{\pm0.1}_{2.3\sigma/0.0\sigma}$ \\[3pt] $2.0^{\pm0.2}_{2.2\sigma/0.0\sigma}$  \\[3pt] $1.7^{\pm0.1}_{2.3\sigma/0.0\sigma}$ \\[3pt] \makecell{$3.1^{\pm0.2}_{2.0\sigma/0.0\sigma}$ \\ \\}} &
	\makecell{$1.3^{\pm0.1}_{1.3\sigma/0.0\sigma}$ \\[3pt] $2.1^{\pm0.2}_{4.6\sigma/1.9\sigma}$ \\[3pt] $2.0^{\pm0.1}_{3.4\sigma/0.0\sigma}$ \\[3pt] \makecell{$1.7^{\pm0.3}_{2.4\sigma/0.0\sigma}$ \\[2pt] $3.3^{\pm0.5}_{2.3\sigma/0.0\sigma}$}} &
	\makecell{$1.3^{\pm0.1}_{1.3\sigma/0.0\sigma}$ \\[3pt] $3.1^{\pm0.2}_{4.0\sigma/0.6\sigma}$ \\[3pt] $2.1^{\pm0.2}_{2.5\sigma/0.0\sigma}$ \\[3pt] \makecell{$1.7^{\pm0.3}_{3.2\sigma/0.0\sigma}$ \\[2pt] $3.4^{\pm0.4}_{3.6\sigma/0.0\sigma}$}} &
	\makecell{$3.4^{\pm0.1}_{2.0\sigma/0.0\sigma}$ \\[3pt] $2.2^{\pm0.2}_{4.0\sigma/0.6\sigma}$ \\[3pt] $3.5^{\pm0.2}_{4.0\sigma/0.6\sigma}$ \\ [3pt] \makecell{$3.4^{\pm0.1}_{2.2\sigma/0.0\sigma}$ \\ \\} } & 
	\makecell{$1.7^{\pm0.4}_{1.2\sigma/0.0\sigma}$ \\[3pt] $2.0^{\pm0.2}_{2.0\sigma/0.0\sigma}$ \\[3pt] $2.0^{\pm0.2}_{3.0\sigma/0.0\sigma}$ \\[2pt] \makecell{$1.7^{\pm0.5}_{2.1\sigma/0.0\sigma}$ \\[2pt] $3.4^{\pm0.7}_{2.2\sigma/0.0\sigma}$}} & 
	\makecell{$1.7^{\pm0.2}_{2.3\sigma/0.0\sigma}$ \\[3pt] $2.6^{\pm0.5}_{2.0\sigma/0.0\sigma}$ \\[3pt] $2.6^{\pm0.4}_{2.0\sigma/0.0\sigma}$ \\ \makecell{$1.7^{\pm0.4}_{3.2\sigma/0.0\sigma}$ \\ \\}} &
	\makecell{3.3$\pm$0.3 \\ 3.6$\pm$0.1 \\ 3.2$\pm$0.1 \\ \makecell{$1.7^{+0.4}_{-0.1}$ \\ \\ }} \\[3pt]
        \hline
        S5 0716+714 & 
	\makecell{{\it Swift}/XRT \\[3pt] UVOT ({\it uvw2}) \\[3pt] UVOT ({\it uvm2}) \\[3pt] UVOT ({\it uvw1}) \\[3pt] V-BAND \\[3pt] R-BAND \\[3pt] OVRO} & 
	\makecell{$3.0^{\pm0.6}_{3.0\sigma/0.0\sigma}$ \\[3pt] $3.4^{\pm0.2}_{1.4\sigma/0.0\sigma}$ \\[3pt] $1.5^{\pm0.2}_{2.3\sigma/0.0\sigma}$ \\[3pt] $1.6^{\pm0.2}_{1.9\sigma/0.0\sigma}$ \\ $2.5^{\pm0.1}_{2.6\sigma/0.0\sigma}$ \\[3pt] $3.5^{\pm0.1}_{1.2\sigma/0.0\sigma}$ \\[3pt] $1.0^{\pm0.1}_{1.3\sigma/0.0\sigma}$} &
	\makecell{$1.8^{\pm0.2}_{3.2\sigma/0.0\sigma}$ \\[3pt] $1.4^{\pm0.1}_{1.3\sigma/0.0\sigma}$ \\[3pt] $1.6^{\pm0.2}_{1.8\sigma/0.0\sigma}$ \\[3pt] $2.4^{\pm0.2}_{2.2\sigma/0.0\sigma}$ \\[3pt] $2.8^{\pm0.1}_{2.5\sigma/0.0\sigma}$ \\[3pt] $3.4^{\pm0.4}_{2.2\sigma/0.0\sigma}$ \\[3pt] $1.2^{\pm0.1}_{3.0\sigma/0.0\sigma}$} &
	\makecell{$1.7^{\pm0.3}_{2.7\sigma/0.0\sigma}$ \\[3pt] $3.4^{\pm0.4}_{1.3\sigma/0.0\sigma}$ \\[3pt] $1.6^{\pm0.2}_{2.1\sigma/0.0\sigma}$ \\[3pt] $1.6^{\pm0.2}_{2.3\sigma/0.0\sigma}$ \\[3pt] $2.5^{\pm0.2}_{3.3\sigma/0.0\sigma}$ \\[3pt] $3.5^{\pm0.5}_{2.3\sigma/0.0\sigma}$ \\[3pt] $1.4^{\pm0.1}_{2.0\sigma/0.0\sigma}$} &
	\makecell{$5.0^{\pm0.2}_{2.0\sigma/0.0\sigma}$ \\ [3pt] $3.4^{\pm0.1}_{2.7\sigma/0.0\sigma}$ \\[3pt] $2.9^{\pm0.2}_{3.0\sigma/0.0\sigma}$ \\[3pt] $2.9^{\pm0.1}_{2.8\sigma/0.0\sigma}$ \\[3pt] $3.3^{\pm0.3}_{3.1\sigma/0.0\sigma}$ \\[3pt] $3.5^{\pm0.3}_{2.0\sigma/0.0\sigma}$ \\[3pt] $2.9^{\pm0.2}_{3.0\sigma/0.0\sigma}$} &
	\makecell{$1.7^{\pm0.5}_{0.9\sigma/0.0\sigma}$ \\[3pt] $1.6^{\pm0.6}_{1.0\sigma/0.0\sigma}$ \\[3pt] $1.6^{\pm0.5}_{1.1\sigma/0.0\sigma}$ \\[3pt] $3.4^{\pm0.8}_{1.1\sigma/0.0\sigma}$ \\[3pt] $2.9^{\pm0.1}_{4.0\sigma/0.6\sigma}$ \\[3pt] $\ddagger$$2.9^{\pm0.1}_{2.0\sigma/0.0\sigma}$ \\[3pt] $4.3^{\pm0.6}_{2.2\sigma/0.0\sigma}$} &
	\makecell{$1.2^{\pm0.2}_{2.4\sigma/0.0\sigma}$ \\[3pt] $1.4^{\pm0.3}_{2.3\sigma/0.0\sigma}$ \\[3pt] $4.0^{\pm0.5}_{0.8\sigma/0.0\sigma}$ \\[3pt] $4^{\pm0.6}_{0.9\sigma/0.0\sigma}$ \\[3pt] $2.7^{\pm0.5}_{3.4\sigma/0.0\sigma}$ \\[3pt] $3.5^{\pm0.1}_{1.3\sigma/0.0\sigma}$ \\[3pt] $3.8^{\pm0.7}_{2.5\sigma/0.0\sigma}$} &
	\makecell{$1.7^{+1.2}_{-0.1}$ \\[3pt] $1.4^{+2.0}_{-0.1}$ \\[3pt] 1.5$\pm$0.1 \\[3pt] 1.5$\pm$0.1 \\ $2.8^{+0.8}_{-0.4}$ \\[3pt] 3.4$\pm$0.2 \\[3pt] $2.5^{+1.5}_{-1}$} \\[3pt]
        \hline
        OJ 014 &
	\makecell{OVRO} &
	\makecell{$5.1^{\pm0.2}_{1.3\sigma/0.0\sigma}$} &
	\makecell{$5.1^{\pm0.2}_{1.3\sigma/0.0\sigma}$} &
	\makecell{$5.3^{\pm0.2}_{1.4\sigma/0.0\sigma}$} &
	\makecell{$5.2^{\pm0.2}_{3.6\sigma/0.0\sigma}$} &
	\makecell{$\ddagger$$5.1^{\pm1.2}_{1.2\sigma}$} &
	\makecell{$5.2^{\pm1.1}_{4.0\sigma/0.6\sigma}$} &
	\makecell{5.5$\pm$0.1} \vspace{0.05cm}\\        
 \hline
 \hline
\end{tabular}%
}
\end{table*}
\begin{table*}
\centering
\caption{Results of the \textit{z}-DCF cross-correlation analysis for PKS 0454$-$234, S5 0716+714 and OJ 014. The table also shows the period inferred from the cross-correlation with the $\gamma$-ray LC and the auto-correlation. These periods are associated with the significance levels of the oscillations (high-level, low-level, see Figure \ref{fig:autocorrelation}). We report one significance when the high and low significance levels are the same. The correlation with the $\gamma$ rays is limited by the 28 days sampling of the \textit{Fermi}-LAT LCs}
\label{tab:cross_correlation_appendix}
{%
\begin{tabular}{c|ccccccc}
\hline
\hline
Association & Wavelength & Correlation & Cross-correlation & Auto-correlation \\
 Name &  & Lag {[}days{]} & Period {[}years{]} & Period {[}years{]} \\
 \hline
 PKS 0454$-$234 & 
	\makecell{\textit{Fermi}-LAT \\ SMARTS-B \\ V-BAND \\ STEWARD-V \\ R-BAND \\ STEWARD-R \\ SMARTS-J \\ SMARTS-K} &
	\makecell{-- \\ $-4.6^{\pm1.5}_{>3.0\sigma}$ \\[2pt] $1.1^{\pm0.4}_{>4.0\sigma}$ \\[2pt] $-5.4^{\pm9.9}_{>4.0\sigma}$ \\[2pt] $12.5^{\pm5}_{>4.0\sigma}$ \\[2pt] $-14.9^{\pm9.5}_{>4.0\sigma}$ \\[2pt] $2.6^{\pm1.0}_{>2.0\sigma}$ \\[2pt] $1.3^{\pm3.3}_{>3.0\sigma}$} &
	\makecell{-- \\ -- \\ $1.5^{\pm0.1}_{(1-3)\sigma}$ \\ -- \\ -- \\ -- \\ -- \\ --} &
	\makecell{$3.5^{\pm0.3}_{3\sigma}$ \\ -- \\ $1.5^{\pm0.2}_{(1-2)\sigma}$ \\ -- \\ -- \\ -- \\ -- \\ --} \\
	\hline
	S5 0716+714 & 
	\makecell{\textit{Fermi}-LAT \\ {\it Swift}/XRT \\ UVOT ({\it uvw2}) \\ UVOT ({\it uvm2}) \\ UVOT ({\it uvw1}) \\ V-BAND \\ R-BAND \\ OVRO} &
	\makecell{--\\ $16.0^{\pm2.0}_{3.0\sigma}$ \\[2pt] $11.4^{\pm0.6}_{\approx4.0\sigma}$ \\[2pt] $11.4^{\pm0.5}_{>4.0\sigma}$ \\[2pt] $11.4^{\pm0.5}_{>4.0\sigma}$ \\[2pt] $4.8^{\pm1.5}_{\approx5.0\sigma}$ \\[2pt] $4.8^{\pm0.3}_{>3.0\sigma}$ \\[2pt] $-13.6^{\pm11.8}_{\approx3.0\sigma}$} &
	\makecell{--\\ -- \\ -- \\ -- \\ -- \\ $2.7^{\pm0.2}_{3\sigma}$ \\ $2.8^{\pm0.2}_{(2-3)\sigma}$ \\ $2.7^{\pm0.3}_{2\sigma}$} &
	\makecell{$2.5^{\pm0.2}_{2\sigma}$ \\ -- \\ -- \\ -- \\ -- \\  $2.8^{\pm0.3}_{2\sigma}$ \\ -- \\ --} \\
	\hline
	OJ 014 &
	\makecell{\textit{Fermi}-LAT \\ CSS \\ OVRO} &
	\makecell{-- \\ $11.9^{\pm9.0}_{\approx3.0\sigma}$ \\[2pt] $-40.4^{\pm3.6}_{>4.0\sigma}$} &
	\makecell{-- \\ $4.5^{\pm0.3}_{2\sigma}$ \\ $4.6^{\pm0.3}_{(3-4)\sigma}$} &
	\makecell{$4^{\pm0.3}_{2\sigma}$ \\ -- \\ $4.8^{\pm0.5}_{2\sigma}$} \\
\hline
\hline
\end{tabular}%
}
\end{table*}
\begin{table*}
\centering
\caption{List of power spectral indices inferred with the \texttt{PSRESP} method for the MWL LCs of the blazars. The uncertainty and the probability of each index are also shown. The `X' denotes that the \texttt{PSRESP} method does not converge to a specific value. The ``success fraction'' indicates the goodness of fit of \texttt{PSRESP} by giving the probability of a model being accepted. The power spectral indices of the $\gamma$ rays are obtained from P22.}
\label{tab:slopes}
{%
\begin{tabular}{c|ccc|cc}
\hline
\hline
Association Name & Wavelength &  PSD Fit & Success Fraction & $\gamma$-ray PSD Index \\
\hline
PG 1553+113 &
	\makecell{{\it Swift}/XRT) \\ UVOT ({\it uvw2}) \\ UVOT ({\it uvm2}) \\ UVOT ({\it uvw1}) \\ V-BAND \\ R-BAND \\ OVRO} &
	\makecell{0.8$\pm$0.6 \\ 1.3$\pm$0.8 \\ 1.2$\pm$0.8 \\ 1.0$\pm$0.7 \\ 1.5$\pm$0.5 \\ 1.5$\pm$0.5 \\ 1.6$\pm$0.4} &
	\makecell{8.5 \\ 93.7 \\ 94.5 \\ 13.5 \\ 44.6 \\ 66.5 \\ 77.2} &
		1.2$\pm$0.4\\
	\hline
	PKS 2155$-$304 &
	\makecell{{\it Swift}/XRT) \\ UVOT ({\it uvw2}) \\ UVOT ({\it uvm2}) \\ UVOT ({\it uvw1}) \\ SMARTS-B \\ V-BAND \\ R-BAND \\ SMARTS-J \\ SMARTS-K} &
	\makecell{0.6$\pm$0.3 \\ 1.3$\pm$0.6 \\ 1.3$\pm$0.5 \\ 1.5$\pm$0.7 \\ 1.2$\pm$0.4 \\ 1.5$\pm$0.7 \\ 1.5$\pm$0.5 \\ 1.4$\pm$0.3 \\ 1.4$\pm$0.2} &
	\makecell{7.8 \\ 71.1 \\ 57.7 \\ 90.1 \\ 27.4 \\ 26.6 \\ 71.6 \\ 5.6 \\ 4.1} &
		1.0$\pm$0.6\\
	\hline
PKS 0454$-$234 & 
	\makecell{V-BAND \\ STEWARD-V \\ R-BAND \\ STEWARD-R} &
	\makecell{1.5$\pm$0.4 \\ 1.5$\pm$0.8 \\ 1.7$\pm$0.6 \\ X} &
	\makecell{60.4 \\ 78.9 \\ 32.8 \\ X} &
		1.3$\pm$0.3\\
	\hline
	S5 0716+714 & 
	\makecell{{\it Swift}/XRT) \\ UVOT ({\it uvw2}) \\ UVOT ({\it uvm2}) \\ UVOT ({\it uvw1}) \\ V-BAND \\ R-BAND \\ OVRO} &
	\makecell{1.4$\pm$0.3 \\ 1.1$\pm$0.6 \\ 1.2$\pm$0.7 \\ 1.0$\pm$0.7 \\ 1.2$\pm$0.6 \\ 1.1$\pm$0.6 \\ 1.2$\pm$0.2} &
	\makecell{13.8 \\ 94.4 \\ 92.3 \\ 84.6 \\ 62.0 \\ 69.8 \\ 20.8} &
		1.1$\pm$0.3\\
	\hline
	OJ 014 &
	\makecell{OVRO} &
	\makecell{1.6$\pm$0.8} &
	\makecell{19.0} &
	1.2$\pm$0.4\\
\hline
\hline
\end{tabular}%
}
\end{table*}
\begin{table*}
\centering
\caption{Results of the fractional variability and the periods inferred from the structure functional (see $\S$\ref{sec:variability}).}
\label{tab:fractional}
{%
\begin{tabular}{c|cccc}
\hline
\hline
Association & Wavelength & $F_{\text{var}}$ & $\tau_{\text{SF}}$ \\
 Name &  &  \\
 \hline
 	PG 1553+113 &
	\makecell{\textit{Fermi}-LAT \\ {\it Swift}/XRT \\ UVOT ({\it uvw2}) \\ UVOT ({\it uvm2}) \\ UVOT ({\it uvw1})  \\ V-BAND \\ R-BAND \\ OVRO} &
		\makecell{0.24$\pm$0.06 \\ 0.43 $\pm$0.08 \\ 0.25$\pm$0.01 \\ 0.25$\pm$0.01 \\ 0.25$\pm$0.01 \\ 0.21$\pm$0.02 \\ 0.22$\pm$0.01 \\ 0.12$\pm$0.02} &
	\makecell{-- \\ 0.4$\pm$2.1 \\ 2.4$\pm$0.1 \\ 2.4$\pm$0.1 \\ 2.4$\pm$0.1 \\ 2.4$\pm$0.1 \\ 2.5$\pm$0.1 \\ 2.4$\pm$0.1}
	\\
	\hline
	PKS 2155$-$304 &
	\makecell{\textit{Fermi}-LAT \\ {\it Swift}/XRT \\ UVOT ({\it uvw2}) \\ UVOT ({\it uvm2}) \\ UVOT ({\it uvw1}) \\ SMARTS-B \\ V-BAND \\ R-BAND \\ SMARTS-J \\ SMARTS-K} &
		\makecell{0.47$\pm$0.04 \\ 0.14$\pm$0.22 \\ 0.37$\pm$0.01 \\ 0.37$\pm$0.01 \\ 0.37$\pm$0.01 \\ 0.37$\pm$0.01 \\ 0.37$\pm$0.01 \\ 0.35$\pm$0.01 \\ 0.37$\pm$0.01 \\ 0.43$\pm$0.01} &
	\makecell{--\\ 1.3$\pm$0.1 \\ 1.3$\pm$0.1 \\ 1.3$\pm$0.1 \\ 1.3$\pm$0.1 \\ 1.8$\pm$0.1 \\ 1.7$\pm$0.1 \\ 2.1$\pm$0.1 \\ 1.8$\pm$0.1 \\ 1.8$\pm$0.1} \\
 	\hline
        PKS 0454$-$234 & 
	\makecell{\textit{Fermi}-LAT \\ SMARTS-B \\ V-BAND \\ STEWARD-V \\ R-BAND \\ STEWARD-R \\ SMARTS-J \\ SMARTS-K} &
        \makecell{0.71$\pm$0.02 \\ 0.75$\pm$0.01 \\ 0.72$\pm$0.01 \\ 0.66$\pm$0.01 \\ 0.75$\pm$0.01 \\ 0.67$\pm$0.01 \\ 0.62$\pm$0.01 \\ 0.62$\pm$0.01} &
	\makecell{-- \\ 1.9$\pm$0.2 \\ 1.8$\pm$0.2 \\ 2.5$\pm$0.2 \\ 3.4$\pm$0.2 \\ 3.4$\pm$0.2 \\ 2.3$\pm$0.2 \\ 2.3$\pm$0.2} \\
    \hline
        S5 0716+714 & 
	\makecell{\textit{Fermi}-LAT \\ {\it Swift}/XRT \\ UVOT ({\it uvw2}) \\ UVOT ({\it uvm2}) \\ UVOT ({\it uvw1}) \\ V-BAND \\ R-BAND \\ OVRO} &
        \makecell{0.57$\pm$0.02 \\ 0.11$\pm$0.22 \\ 0.68$\pm$0.02 \\ 0.68$\pm$0.02 \\ 0.68$\pm$0.02 \\ 0.51$\pm$0.02 \\ 0.53$\pm$0.02 \\ 0.34$\pm$0.01} &
	\makecell{-- \\ 3.3$\pm$0.2 \\ 2.9$\pm$0.2 \\ 2.9$\pm$0.2 \\ 2.9$\pm$0.2 \\ 2.5$\pm$0.2 \\ 2.4$\pm$0.2 \\ 2.9$\pm$0.2} \\
    \hline
    OJ 014 &
	\makecell{\textit{Fermi}-LAT \\ CSS \\ OVRO} &
        \makecell{0.59$\pm$0.06 \\ 0.55$\pm$0.01 \\ 0.38$\pm$0.02} &
	\makecell{-- \\ 1.2$\pm$0.1 \\ 5.2$\pm$0.1} \\
\hline
\hline
\end{tabular}%
}
\end{table*}

\subsection {Software}
\begin{enumerate}
	\item astropy \citep{astropy_2013, astropy_2018}, 
	\item emcee \citep {emcee}, 
	\item \texttt{Fitdistrplus} \citep{delignette_2015}, 
	\item \texttt{PSRESP} (\url{https://github.com/wegenmat-privat/psresp}), 
	\item \texttt{Statsmodels} (\url{https://www.statsmodels.org/stable/index.html}), 
	\item SciPy \citep {SciPy},
	\item Simulating light curves \citep{connolly_code},
	\item z-DFC \citep{zdfc_alexander}.
\end{enumerate}

\subsection {Figures} \label{sec:plots}
\clearpage
\begin{figure*}
	\centering
	\includegraphics[scale=0.25]{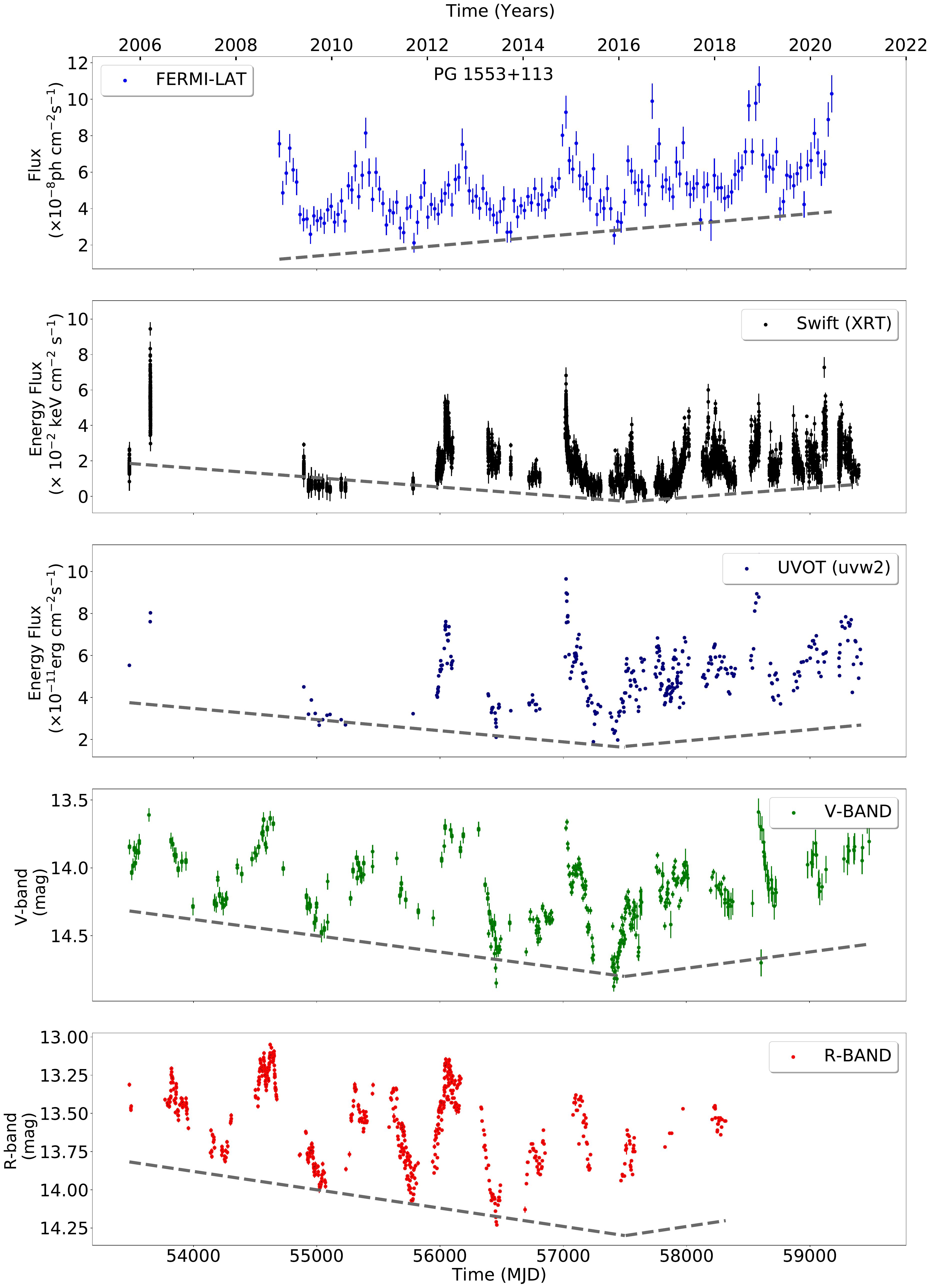}
	\caption{Multiwavelength light curves of PG~1553+113. From top to bottom: \textit{Fermi}-LAT (E $>$ 0.1 GeV), {\it Swift}/XRT), UVOT (filter 'uvw2'), V-band (combination of CSS and ASAS-SN), R-band (combination of Tuorla and KAIT) light curves. The dashed grey line denotes the increasing/decreasing trend.}
	\label{fig:lc_pg_1553}
\end{figure*}

\begin{figure*}
	\centering
	\includegraphics[scale=0.25]{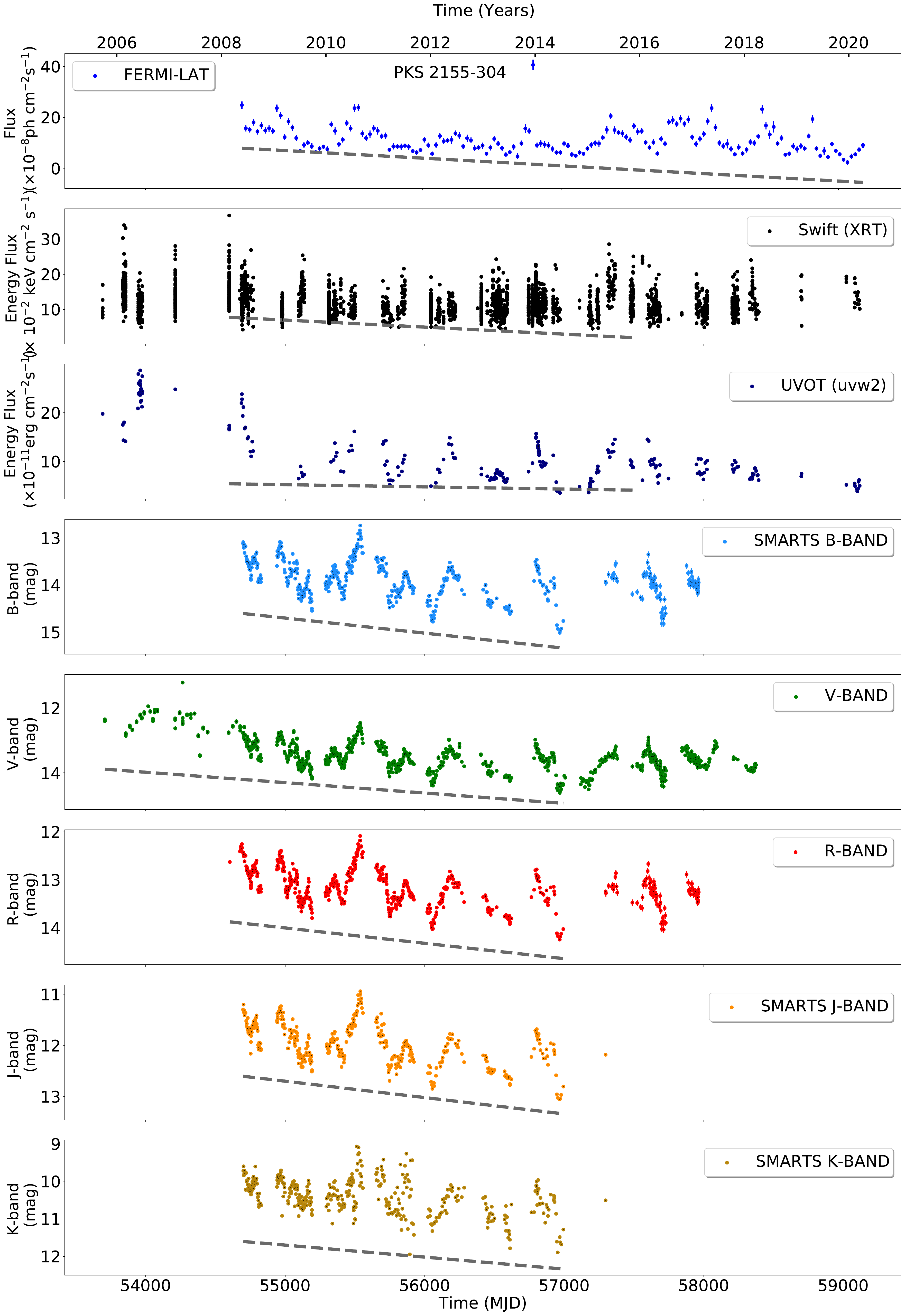}
	\caption{MWL light curves of PKS 2155$-$304. From top to bottom: \textit{Fermi}-LAT (E $>$ 0.1 GeV), {\it Swift}/XRT), UVOT (filter 'uvw2'), B-band (SMARTS), V-band (combination of CSS, SMARTS, and ASAS-SN), R-band (combination of SMARTS, and Tuorla), J-band (SMARTS), and K-band (SMARTS)light curves. The dashed grey line denotes the decreasing trend.\label{fig:mwl_pks2155}}
\end{figure*}

\begin{figure*}
	\centering
	\includegraphics[scale=0.25]{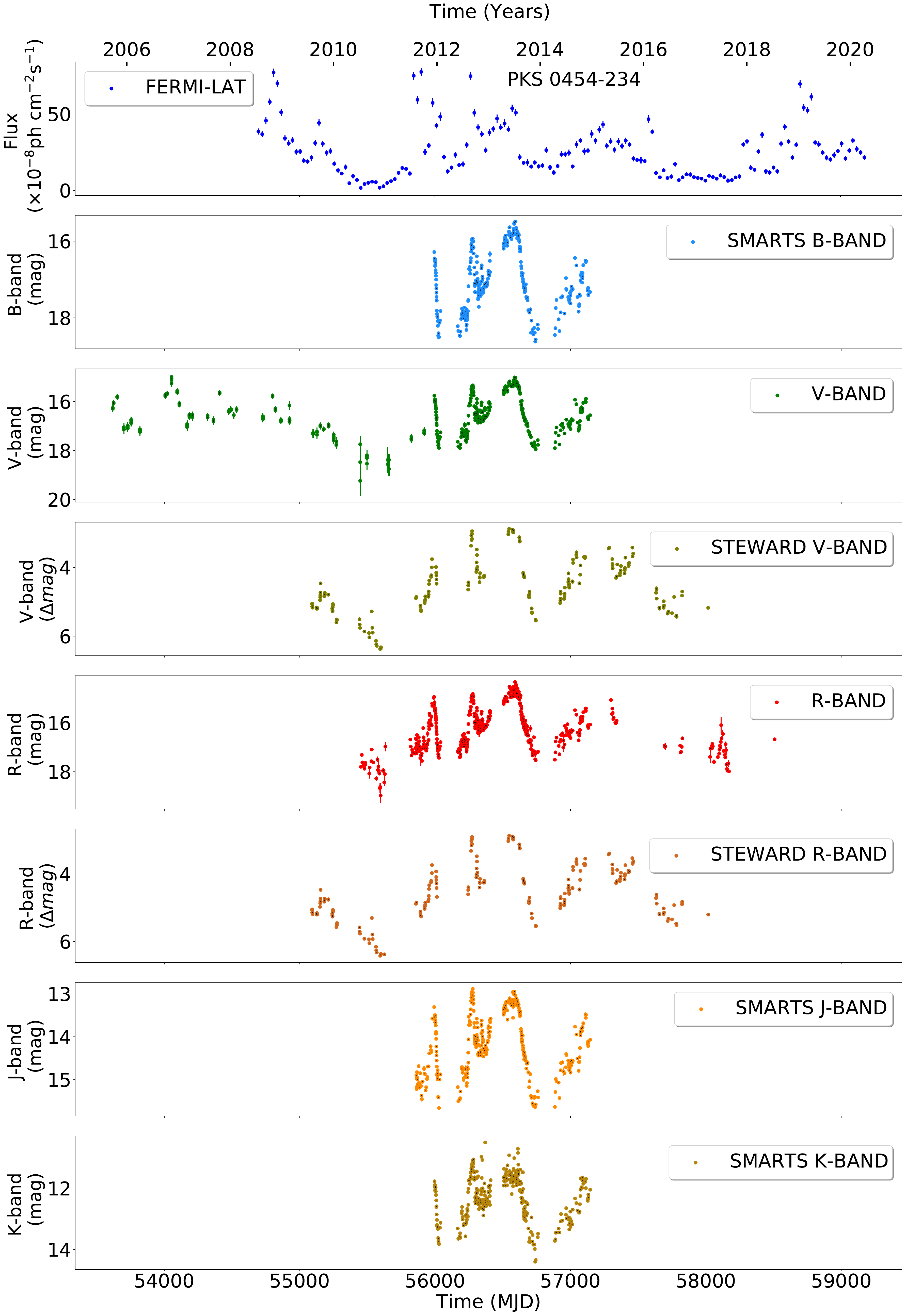}
	\caption{MWL light curves of PKS 0454$-$234. From top to bottom: \textit{Fermi}-LAT (E $>$ 0.1 GeV), B-Band (SMARTS), V-band (combination of SMARTS and CSS), non-calibrated V-band (Steward Observatory), polarized V-band (Steward Observatory), R-band (combination of KAIT and SMARTS), non-calibrated R-band (Steward Observatory), J-band (SMARTS), and K-band (SMARTS) light curves.\label{fig:mwl_pks0454}}
\end{figure*}

\begin{figure*}
	\centering
	\includegraphics[scale=0.25]{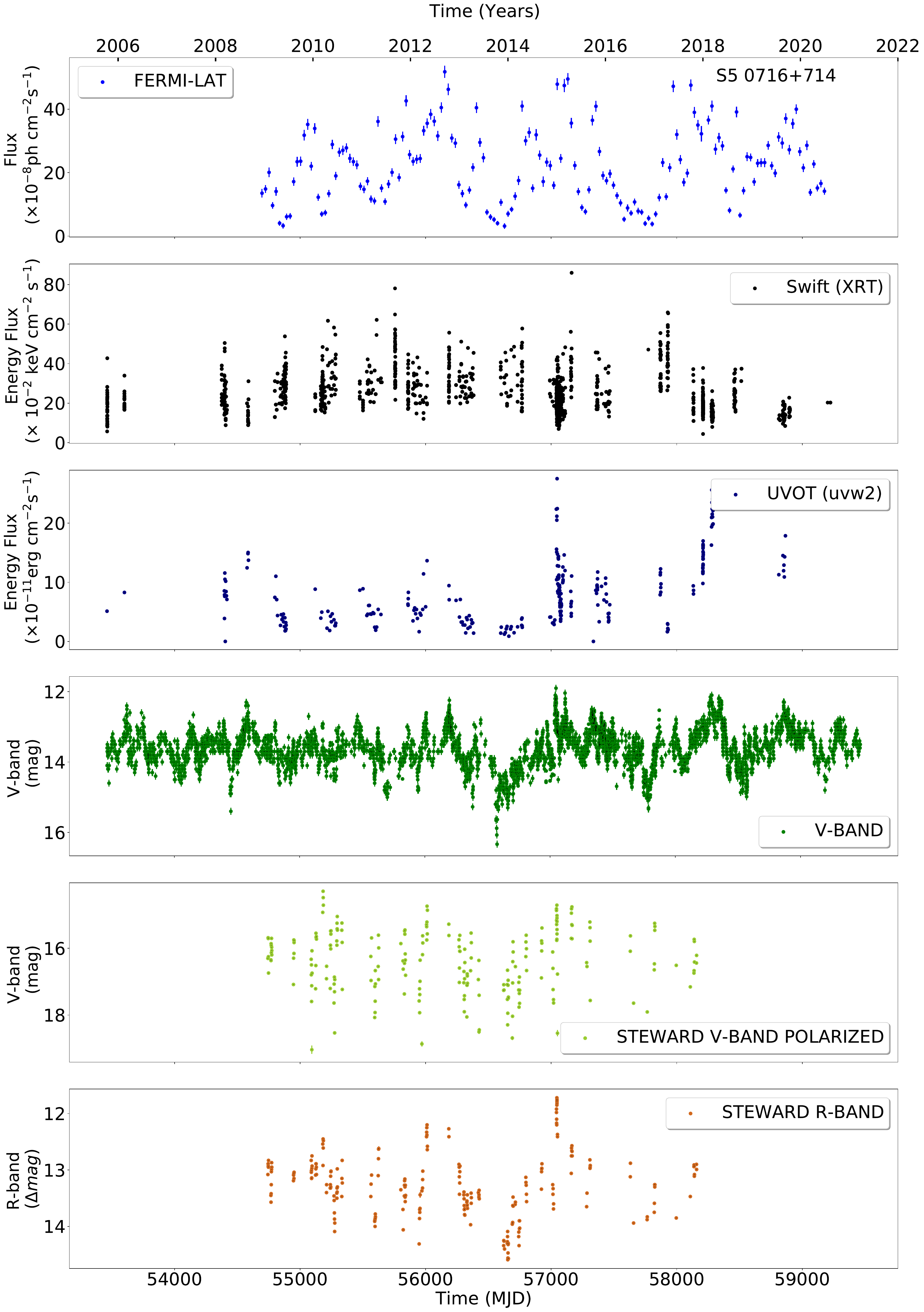}
	\caption{MWL light curves of S5 0716+714. From top to bottom: \textit{Fermi}-LAT (E $>$ 0.1 GeV), {\it Swift}$\backslash$XRT), UVOT (filter 'uvw2'), V-band (combination of AAVSO, ASAS-SN, and Steward Observatory), polarized V-band (Steward Observatory), R-band light (Steward Observatory) curves.\label{fig:mwl_s50716}}
\end{figure*}

\begin{figure*}
	\centering
	\includegraphics[scale=0.32]{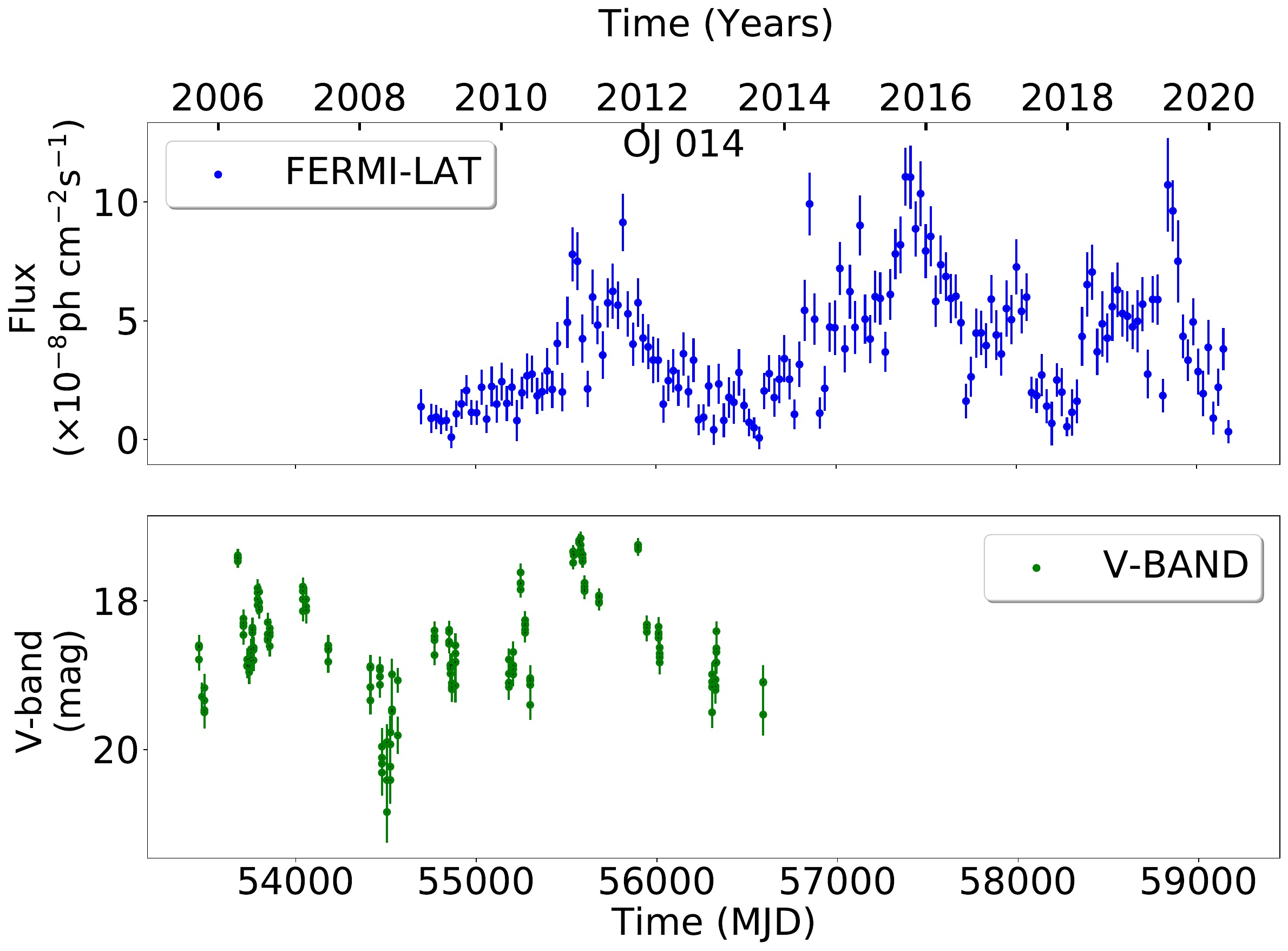}
	\caption{MWL light curves of OJ 014. From top to bottom: \textit{Fermi}-LAT (E $>$ 0.1 GeV), and V-band (combination of CSS and ASAS-SN) light curves. \label{fig:mwl_oj014}}
\end{figure*}

\bsp	
\label{lastpage}
\end{document}